\documentclass[preprint,12pt,a4paper]{elsarticle}

\makeatletter
\def\ps@pprintTitle{%
  \let\@oddhead\@empty
  \let\@evenhead\@empty
  \def\@oddfoot{\reset@font\hfil\thepage\hfil}
  \let\@evenfoot\@oddfoot
}
\makeatother

\usepackage{amssymb}

\usepackage[fleqn]{amsmath}
\begin{document}

\begin{frontmatter}




\title{Turbulence intensity and the friction factor for smooth- and rough-wall pipe flow}


\author[au1]{Nils T. Basse}
\ead{nils.basse@npb.dk}

\address[au1]{Tofteh\o j 23, H\o ruphav, 6470 Sydals, Denmark \\ \vspace{10 mm} \small {\rm \today}}

\begin{abstract}
Turbulence intensity profiles are compared for smooth- and rough-wall pipe flow measurements made in the Princeton Superpipe. The profile development in the transition from hydraulically smooth to fully rough flow displays a propagating sequence from the pipe wall towards the pipe axis. The scaling of turbulence intensity with Reynolds number shows that the smooth- and rough wall level deviates with increasing Reynolds number. We quantify the correspondence between turbulence intensity and the friction factor.

\end{abstract}


\end{frontmatter}


\section{Introduction}
\label{sec:intro}

Measurements of streamwise turbulence in smooth and rough pipes have been carried out in the Princeton Superpipe \cite{hultmark_a} \cite{hultmark_b} \cite{smits_a}. We have treated the smooth pipe measurements as a part of \cite{russo_a}. In this paper, we add the rough pipe measurements to our previous analysis. The smooth (rough) pipe had a radius $R$ of 64.68 (64.92) mm and an RMS roughness of 0.15 (5) $\mu$m, respectively. The corresponding sand-grain roughness is 0.45 (8) $\mu$m \cite{langelandsvik_a}.

The smooth pipe is hydraulically smooth for all Reynolds numbers $Re$ covered. The rough pipe evolves from hydraulically smooth through transitionally rough to fully rough with increasing $Re$. Throughout this paper, $Re$ means the bulk $Re$ defined  using the pipe diameter $D$.

We define the turbulence intensity (TI) $I$ as:

\begin{equation}
I(r) = \frac{v_{\rm RMS}(r)}{v(r)},
\end{equation}

\noindent where $v$ is the mean flow velocity, $v_{\rm RMS}$ is the RMS of the turbulent velocity fluctuations and $r$ is the radius ($r=0$ is the pipe axis, $r=R$ is the pipe wall).

The aim of this paper is to provide the fluid mechanics community with a scaling of the TI with $Re$, both for smooth- and rough-wall pipe flow. An application example is computational fluid dynamics (CFD) simulations where the TI at an opening can be specified. A scaling expression of TI with $Re$ is provided as Eq. (6.62) in \cite{ansys_a}. However, this formula does not appear to be documented, i.e. no reference is provided.

Our paper is structured as follows: In Section \ref{sec:ti_profs}, we study how the TI profiles change over the transition from smooth to rough pipe flow. Thereafter we present the resulting scaling of the TI with $Re$ in Section \ref{sec:ti_scaling}. Quantification of the correspondence between the friction factor and the TI is contained in Section \ref{sec:ff_ti} and we discuss our findings in Section \ref{sec:disc}. Finally, we conclude in Section \ref{sec:conc}.

\section{Turbulence intensity profiles}
\label{sec:ti_profs}

\begin{figure}[htbp]
\includegraphics[width=7cm]{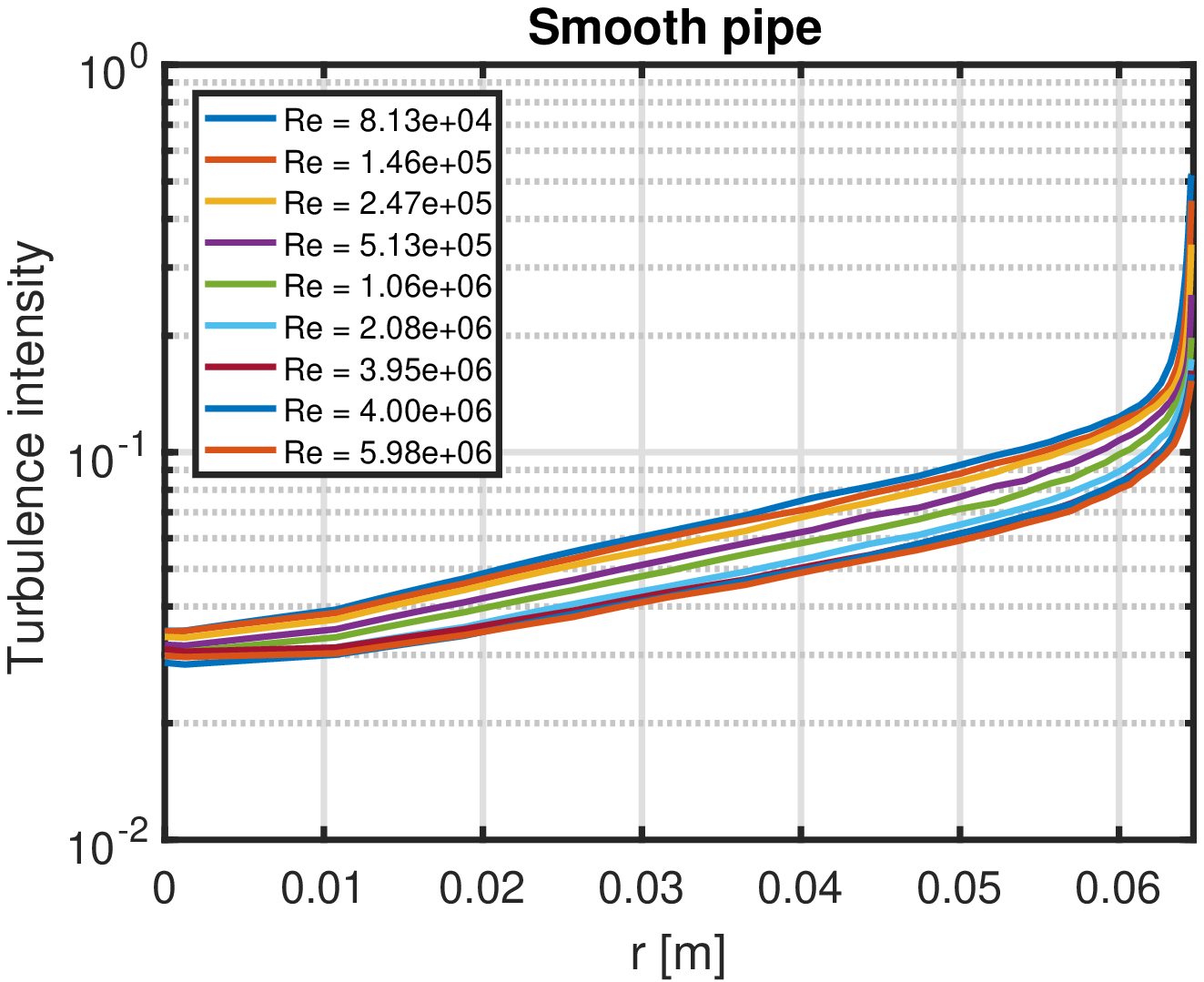}
\hspace{0.5cm}
\includegraphics[width=7cm]{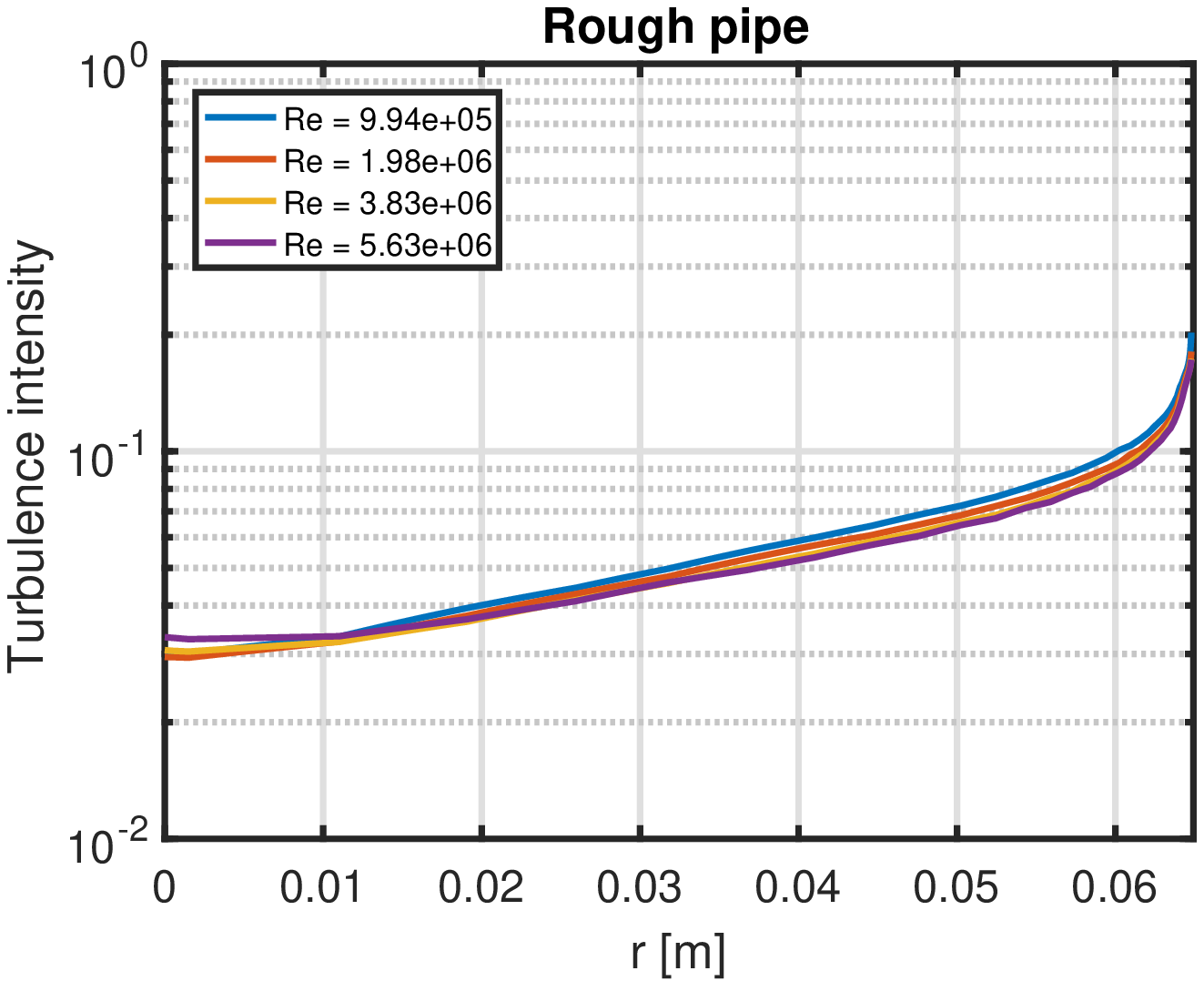}
\caption{Turbulence intensity as a function of pipe radius, left: Smooth pipe, right: Rough pipe.}
\label{fig:meas_TI_smooth_rough}
\end{figure}

We have constructed the TI profiles for the measurements available, see Fig. \ref{fig:meas_TI_smooth_rough}. Nine profiles are available for the smooth pipe and four for the rough pipe. In terms of $Re$, the rough pipe measurements are a subset of the smooth pipe measurements.

\begin{figure}[htbp]
\includegraphics[width=7cm]{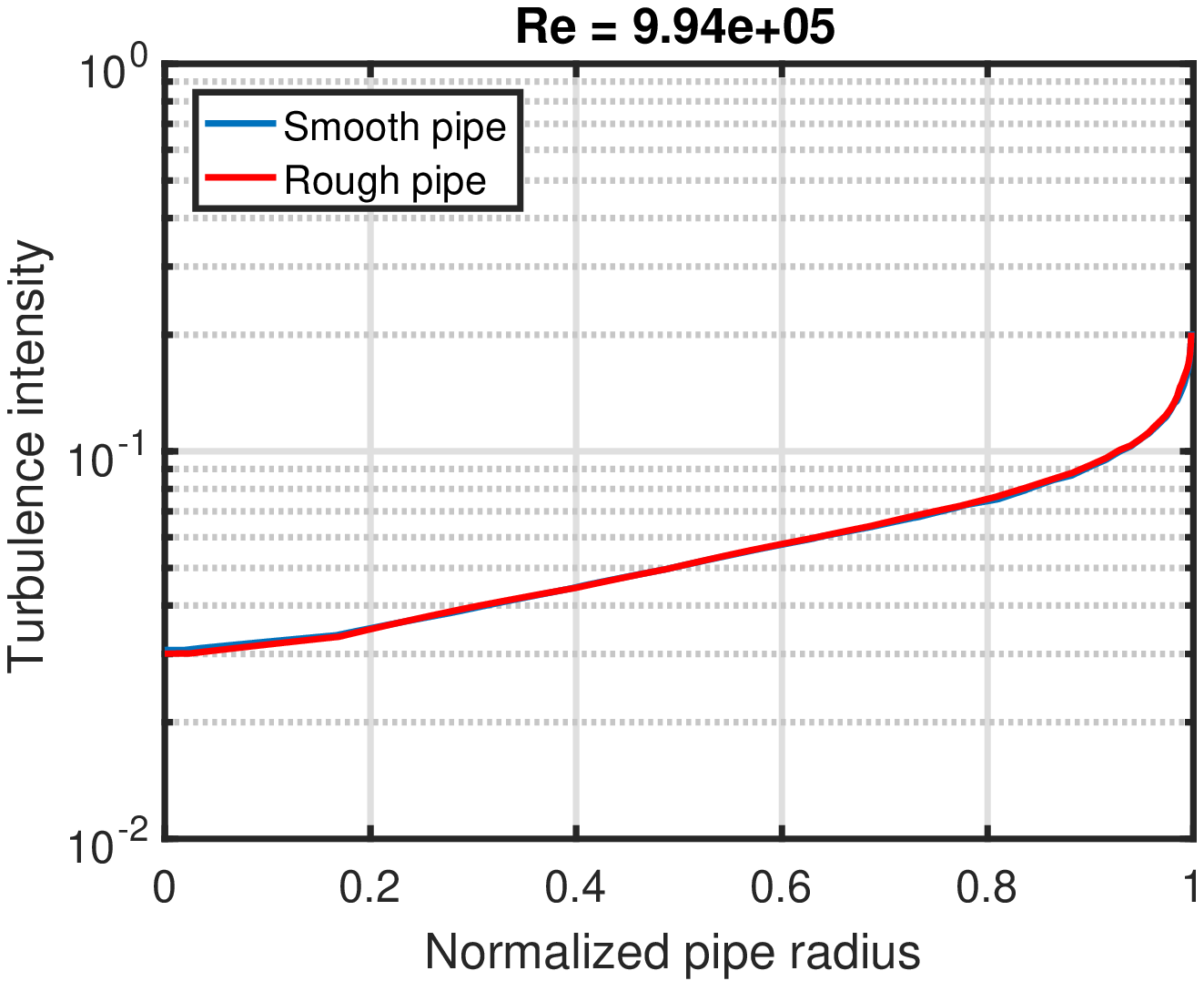}
\hspace{0.5cm}
\includegraphics[width=7cm]{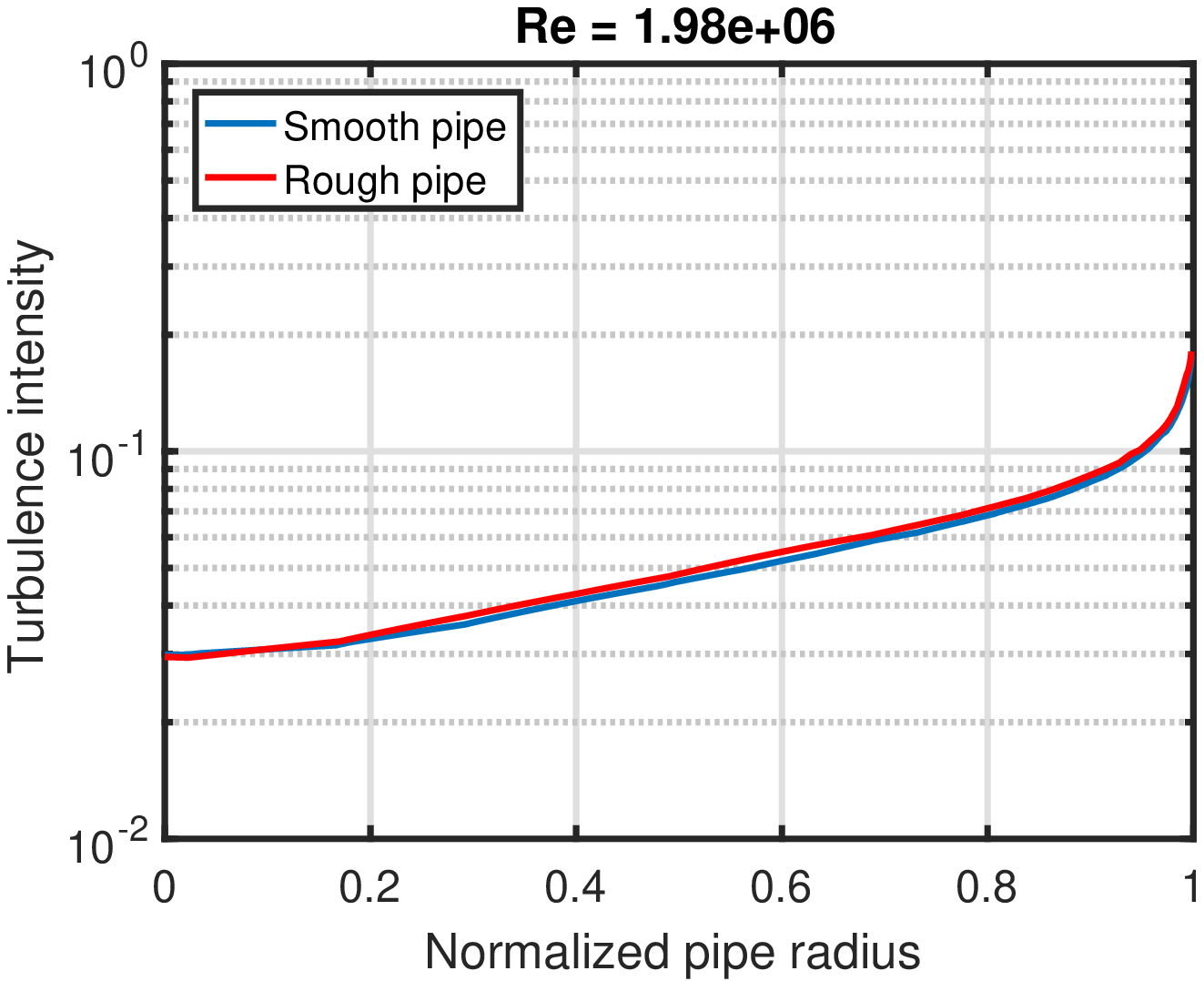}

\vspace{0.5cm}
\includegraphics[width=7cm]{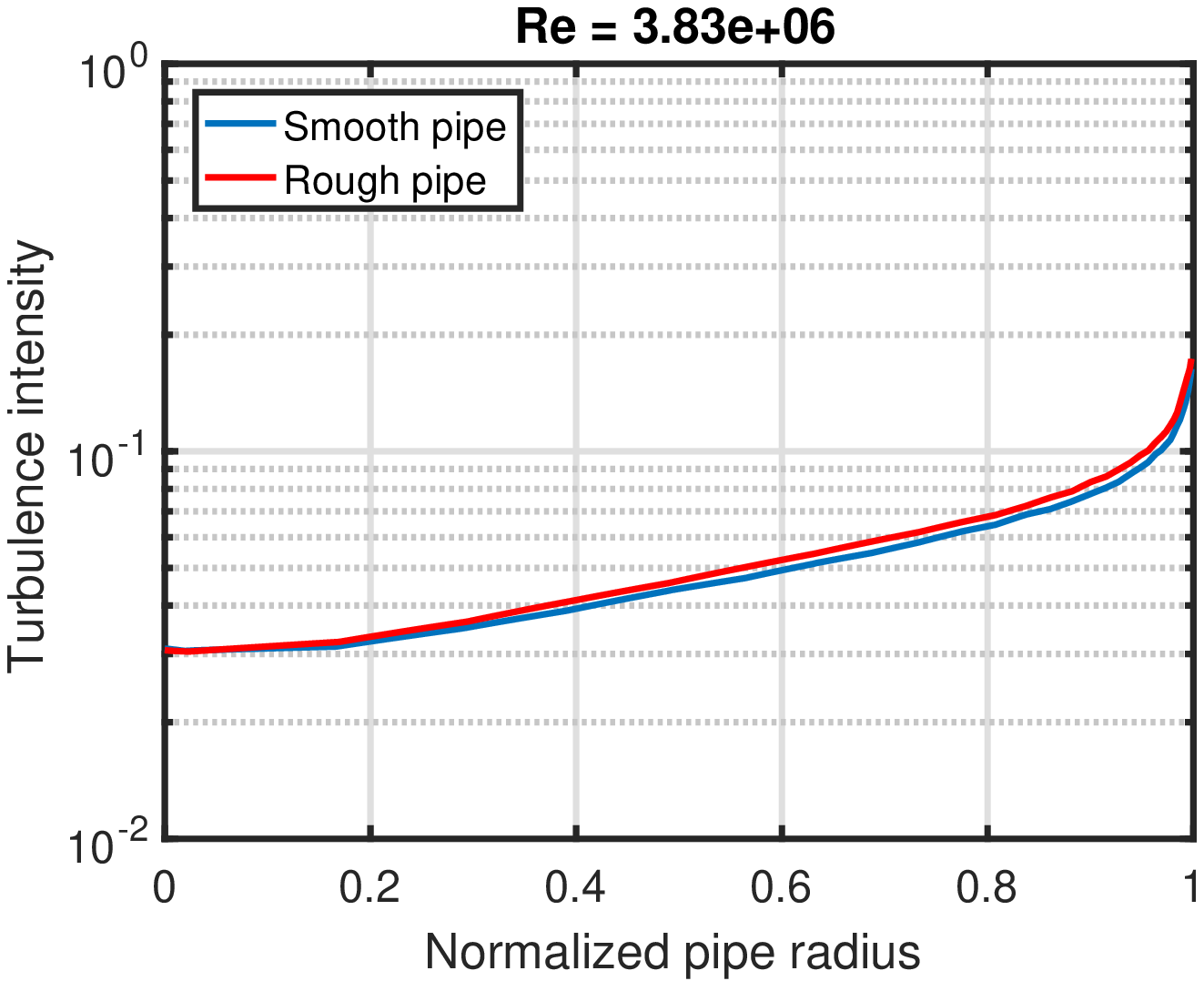}
\hspace{0.5cm}
\includegraphics[width=7cm]{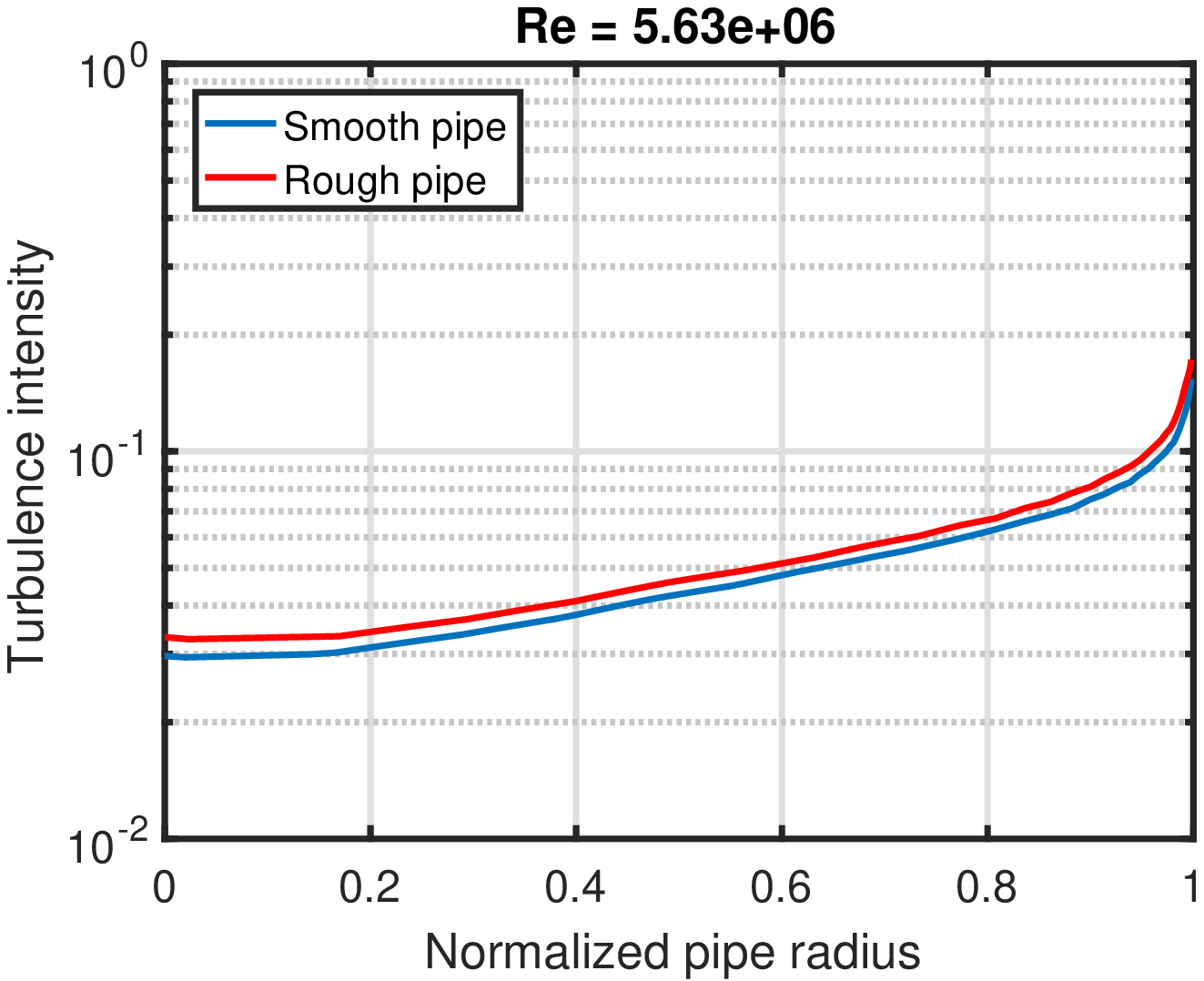}
\caption{Comparison of smooth and rough pipe TI profiles for the four $Re$ values where the rough pipe measurements are done.}
\label{fig:meas_TI_series}
\end{figure}

To make a direct comparison of the smooth and rough pipe measurements, we interpolate the smooth pipe measurements to the four $Re$ values where the rough pipe measurements are done. Further, we use a normalized pipe radius $r_n = r/R$ to account for the difference in smooth and rough pipe radii. The result is a comparison of the TI profiles at four $Re$, see Fig. \ref{fig:meas_TI_series}. As $Re$ increases, we observe that the rough pipe TI becomes larger than the smooth pipe TI.

To make the comparison more quantitative, we define the turbulence intensity ratio (TIR):

\begin{equation}
r_{I, {\rm Rough/Smooth}}(r_n)=\frac{I_{\rm Rough}(r_n)}{I_{\rm Smooth}(r_n)}=\frac{v_{\rm RMS, Rough}(r_n)}{v_{\rm RMS, Smooth}(r_n)} \times \frac{v_{\rm Smooth}(r_n)}{v_{\rm Rough}(r_n)}
\end{equation}

\begin{figure}[htbp]
\includegraphics[width=7cm]{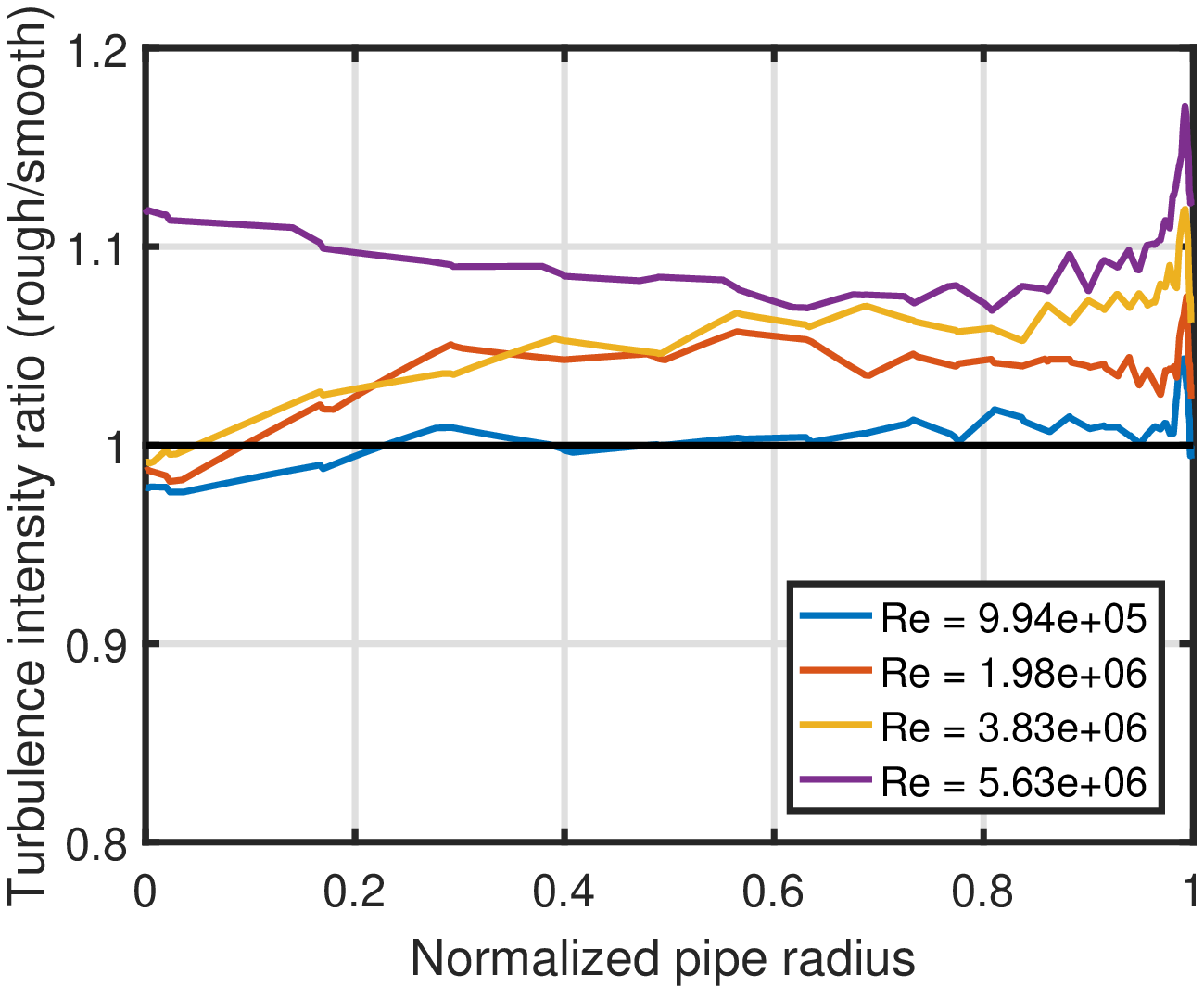}
\hspace{0.5cm}
\includegraphics[width=7cm]{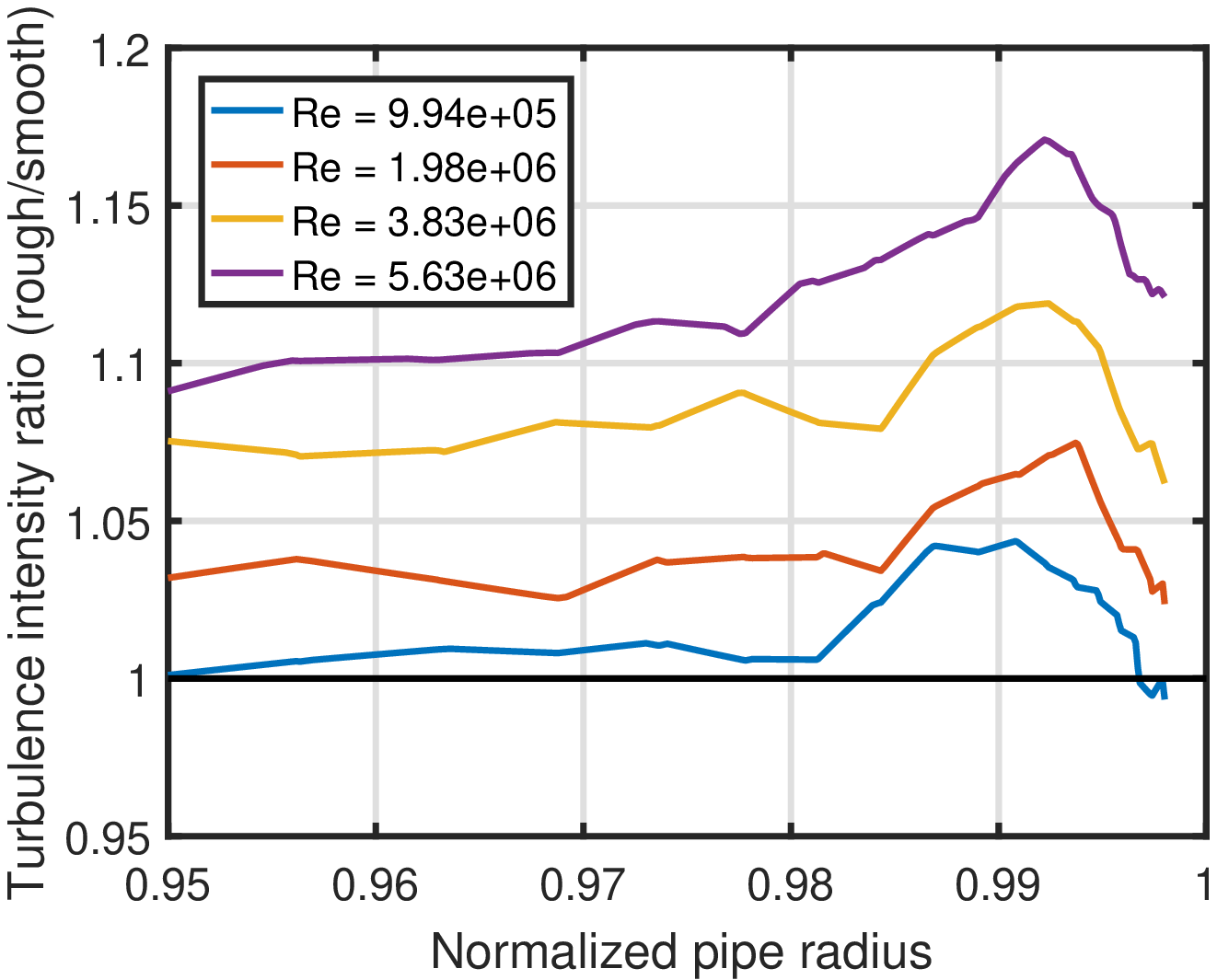}
\caption{Turbulence intensity ratio, left: All radii, right: Zoom to outer 5\%.}
\label{fig:ratio_TI_smooth_rough}
\end{figure}

The TIR is shown in Fig. \ref{fig:ratio_TI_smooth_rough}. The left-hand plot shows all radii; prominent features are:

\begin{itemize}
\item The TIR on the axis is roughly one except for the highest $Re$ where it exceeds 1.1.
\item In the intermediate region between the axis and the wall, an increase is already visible for the second-lowest $Re$, 1.98 $\times$ 10$^6$.
\end{itemize}

The events close to the wall are most clearly seen in the right-hand plot of Fig. \ref{fig:ratio_TI_smooth_rough}. A local peak of TIR is observed for all $Re$; the magnitude of the peak increases with $Re$. Note that we only analyse data to 99.8\% of the pipe radius. So the 0.13 mm closest to the wall are not considered.

\begin{figure}[htbp]
\includegraphics[width=14cm]{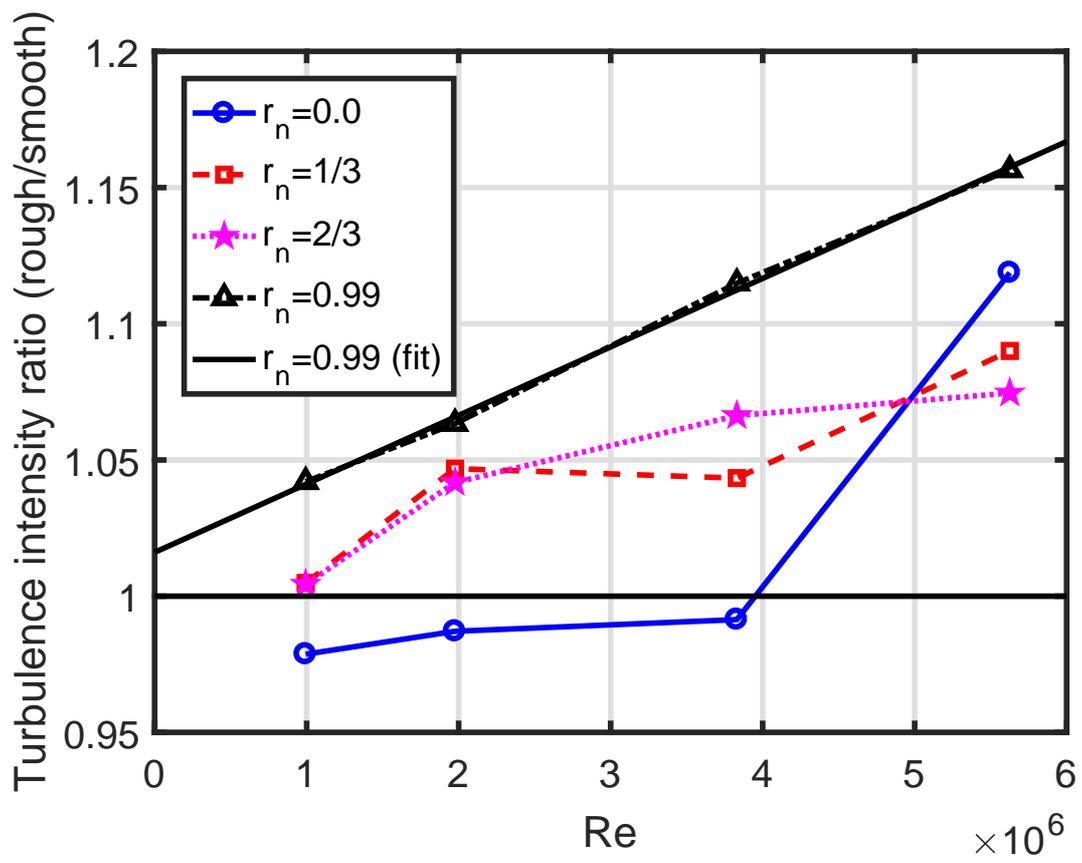}
\caption{Turbulence intensity ratios for fixed $r_n$.}
\label{fig:rn_RE_TI}
\end{figure}

The TIR information can also be represented by studying the TIR at fixed $r_n$ vs. $Re$, see Fig. \ref{fig:rn_RE_TI}. From this plot we find that the magnitude of the peak close to the wall ($r_n = 0.99$) increases linearly with $Re$:

\begin{equation}
r_{I, {\rm Rough/Smooth}}(r_n=0.99)=2.5137 \times 10^{-8} \times Re + 1.0161
\end{equation}

Information on fits of the TI profiles to analytical expressions can be found in \ref{app:ti_prof_fits}.

\section{Turbulence intensity scaling}
\label{sec:ti_scaling}

We define the TI averaged over the pipe area as:

\begin{equation}
I_{\rm Pipe~area} = \frac{2}{R^2} \int_0^R \frac{v_{\rm RMS}(r)}{v(r)} r {\rm d}r
\label{eq:ti_area_AD}
\end{equation}

In \cite{russo_a}, another definition was used for the TI averaged over the pipe area. Analysis presented in Sections \ref{sec:ti_scaling} and \ref{sec:ff_ti} is repeated using that definition in \ref{app:alt_def_ti}.

\begin{figure}[htbp]
\includegraphics[width=14cm]{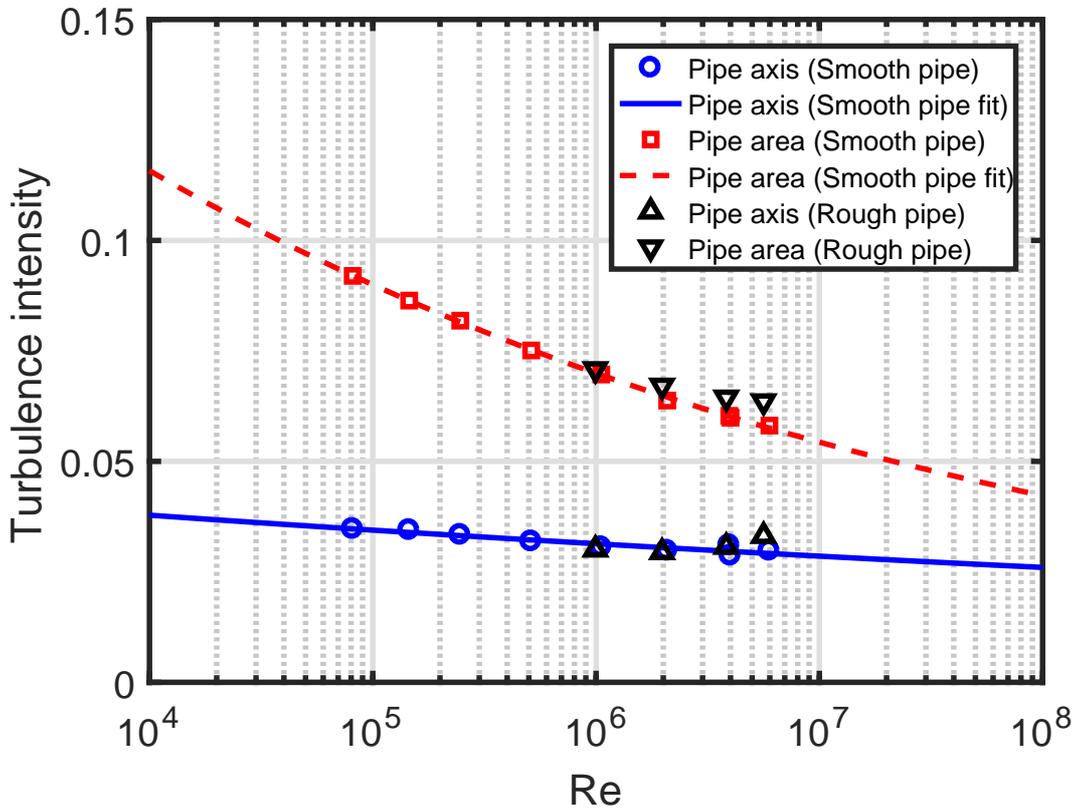}
\caption{Turbulence intensity for smooth and rough pipe flow.}
\label{fig:scal_smooth_rough}
\end{figure}

Scaling of the TI with $Re$ for smooth- and rough-wall pipe flow is shown in Fig. \ref{fig:scal_smooth_rough}.

For $Re=10^6$, the smooth and rough pipe values are almost the same. However, when $Re$ increases, the TI of the rough pipe increases compared to the smooth pipe; this increase is to a large extent caused by the TI increase in the intermediate region between the pipe axis and the pipe wall, see Figs. \ref{fig:ratio_TI_smooth_rough} and \ref{fig:rn_RE_TI}. We have not made fits to the rough wall pipe measurements because of the limited number of datapoints.

\vspace{0.5cm}

\section{Friction factor}
\label{sec:ff_ti}

The fits shown in Fig. \ref{fig:scal_smooth_rough} are:

\begin{equation}
\begin{array}{lcl}
I_{\rm Smooth~pipe~axis} &=& 0.0550 \times Re^{-0.0407} \\
I_{\rm Smooth~pipe~area} &=& 0.317 \times Re^{-0.110}
\end{array}
\label{eq:ti_scal}
\end{equation}

The Blasius smooth pipe (Darcy) friction factor \cite{blasius_a} is also expressed as an $Re$ power-law:

\begin{equation}
\lambda_{\rm Blasius} = 0.3164 \times Re^{-0.25}
\label{eq:blasius_scal}
\end{equation}

\begin{figure}[htbp]
\includegraphics[width=14cm]{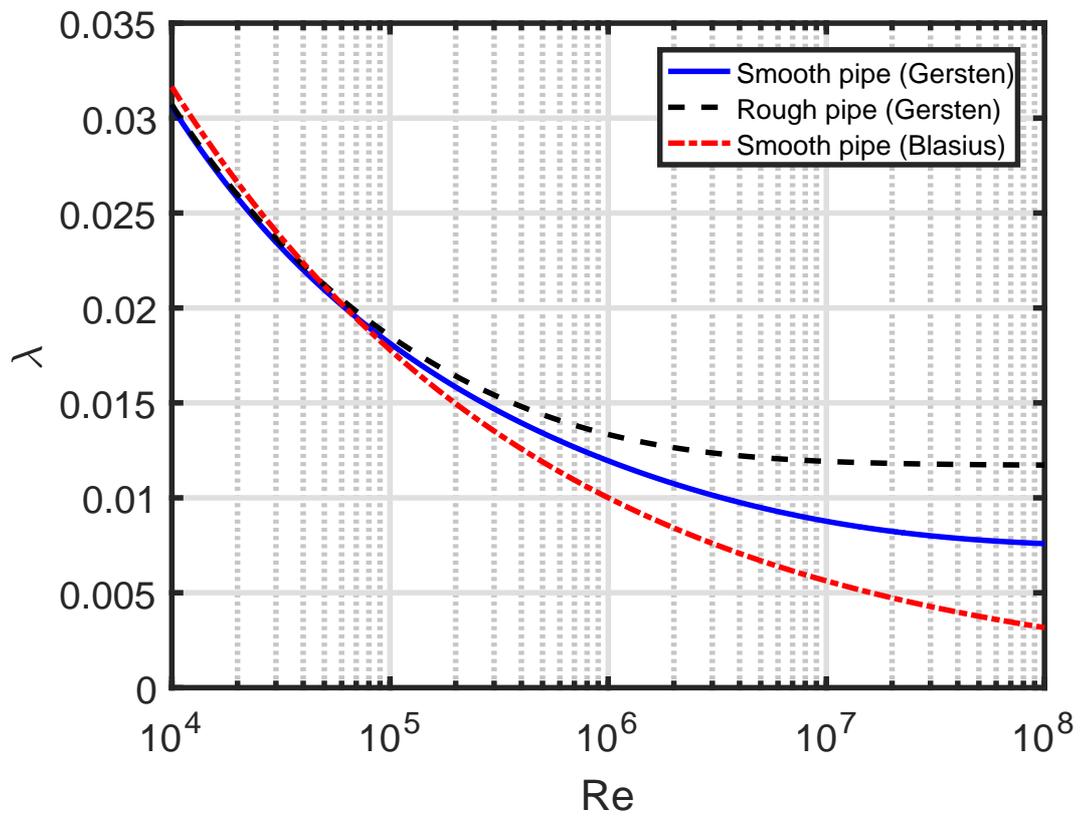}
\caption{Friction factor.}
\label{fig:fric_fac}
\end{figure}

The Blasius friction factor matches measurements best for $Re<10^5$; the friction factor by e.g. Gersten (Eq. (1.77) in \cite{gersten_a}) is preferable for larger $Re$. The Blasius and Gersten friction factors are compared in Fig. \ref{fig:fric_fac}. The deviation between the smooth and rough pipe Gersten friction factors above $Re=10^5$ is qualitatively similar to the deviation between the smooth and rough pipe area TI in Fig. \ref{fig:scal_smooth_rough}. For the Gersten friction factors, we have used the measured pipe roughnesses.

\vspace{0.5cm}

For the smooth pipe, we can combine Eqs. (\ref{eq:ti_scal}) and (\ref{eq:blasius_scal}) to relate the pipe area TI to the Blasius friction factor:

\begin{equation}
\begin{array}{lcl}
I_{\rm Smooth~pipe~area} &=& 0.526 \times \lambda_{\rm Blasius}^{0.44} \\
\lambda_{\rm Blasius} &=&  4.307 \times I_{\rm Smooth~pipe~area}^{2.27}
\end{array}
\label{eq:blasius_ti_rel}
\end{equation}

\begin{figure}[htbp]
\includegraphics[width=14cm]{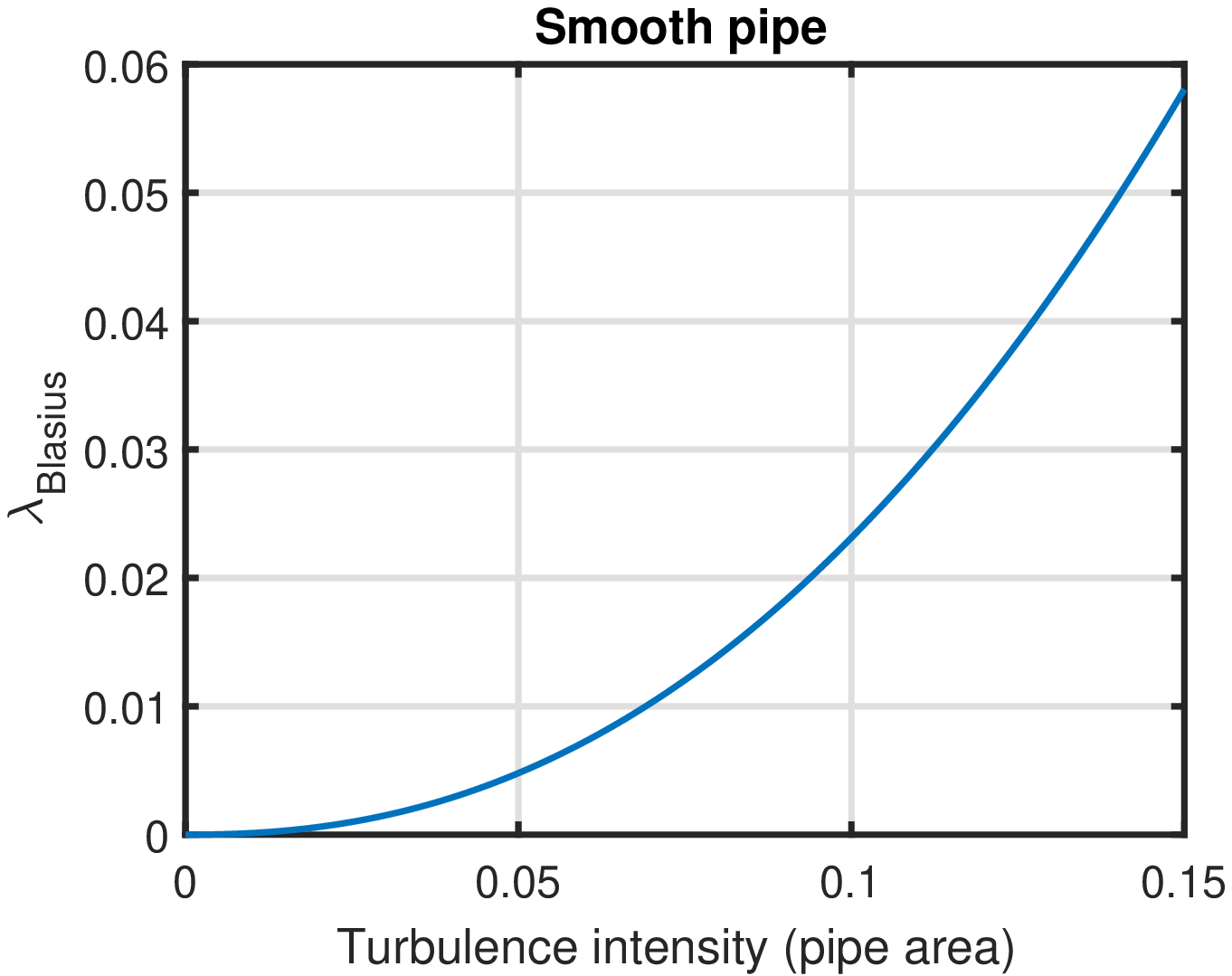}
\caption{Relationship between pipe area turbulence intensity and the Blasius friction factor.}
\label{fig:NEW_ti_fric}
\end{figure}

The TI and Blasius friction factor scaling is shown in Fig. \ref{fig:NEW_ti_fric}.

For axisymmetric flow in the streamwise direction, the mean flow velocity averaged over the pipe area is:

\begin{equation}
v_m = \frac{2}{R^2} \times \int_0^R v(r) r {\rm d}r
\label{eq:avrg_v_m}
\end{equation}

Now we are in a position to define an average velocity of the turbulent fluctuations:

\begin{equation}
\langle v_{\rm RMS} \rangle = v_m I_{\rm Pipe~area} = \frac{4}{R^4} \int_0^R v(r) r {\rm d}r \int_0^R \frac{v_{\rm RMS}(r)}{v(r)} r {\rm d}r
\label{eq:avrg_v_rms}
\end{equation}

The friction velocity is:

\begin{equation}
v_{\tau}=\sqrt{\tau_w/\rho},
\end{equation}

\noindent where $\tau_w$ is the wall shear stress and $\rho$ is the fluid density.

\begin{figure}[htbp]
\includegraphics[width=14cm]{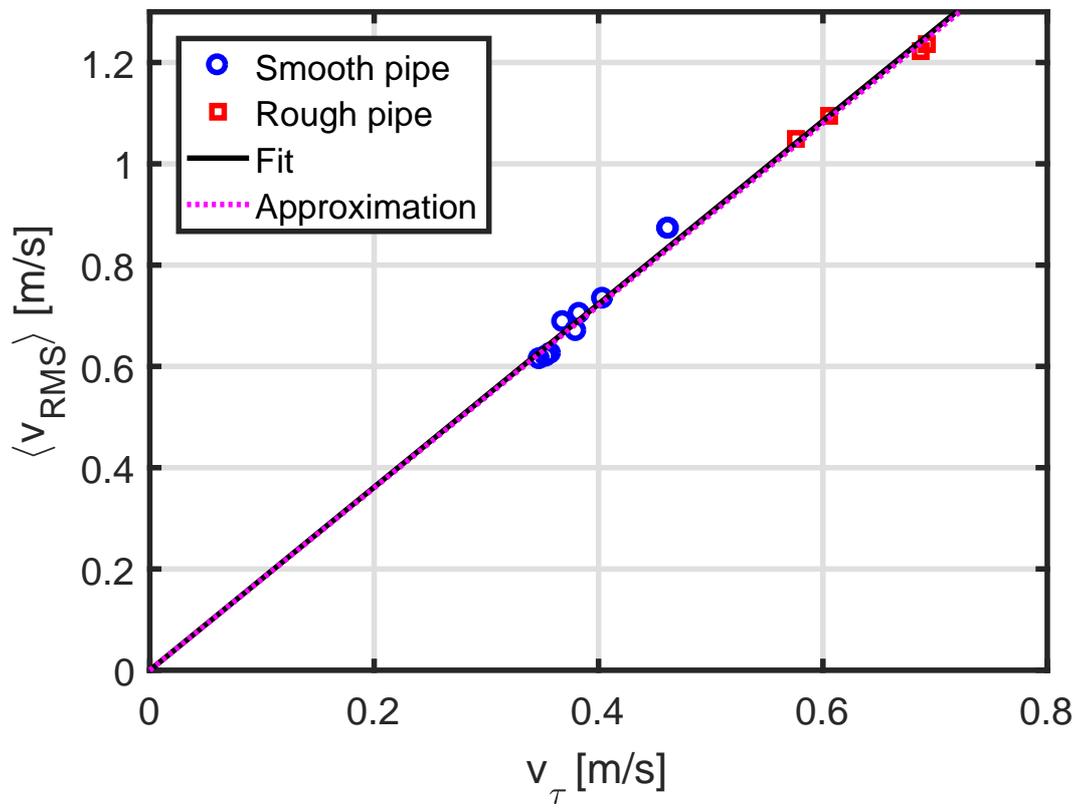}
\caption{Relationship between friction velocity and the average velocity of the turbulent fluctuations.}
\label{fig:fluc_fric}
\end{figure}

The relationship between $\langle v_{\rm RMS} \rangle$ and $v_{\tau}$ is illustrated in Fig. \ref{fig:fluc_fric}. From the fit, we have:

\begin{equation}
\langle v_{\rm RMS} \rangle = 1.8079 \times v_{\tau},
\label{eq:rms_fric_fit}
\end{equation}

\noindent which we approximate as:

\begin{equation}
\langle v_{\rm RMS} \rangle \sim \frac{9}{5} \times v_{\tau}
\label{eq:rms_fric_approx}
\end{equation}

Eqs. (\ref{eq:rms_fric_fit}) and (\ref{eq:rms_fric_approx}) above correspond to the usage of the friction velocity as a proxy for the velocity of the turbulent fluctuations \cite{schlichting_a}. We note that the rough wall velocities are higher than for the smooth wall.

\vspace{0.5cm}

Eqs. (\ref{eq:avrg_v_rms}) and (\ref{eq:rms_fric_approx}) can be combined with Eq. (1.1) in \cite{mckeon_a}:

\begin{equation}
\lambda = \frac{4 \tau_w}{\frac{1}{2} \rho v_m^2} = \frac{- \left( \Delta P/L \right) D}{\frac{1}{2} \rho v_m^2} = 8 \times \frac{v_{\tau}^2}{v_m^2} \sim \frac{200}{81} \times I_{\rm Pipe~area}^2,
\label{eq:d_w}
\end{equation}

\noindent where $\Delta P$ is the pressure loss, $L$ is the pipe length and $D$ is the pipe diameter. This can be reformulated as:

\begin{equation}
I_{\rm Pipe~area} \sim \frac{9}{10 \sqrt{2}} \times \sqrt{\lambda}
\label{eq:ti_ff}
\end{equation}

\begin{figure}[htbp]
\includegraphics[width=14cm]{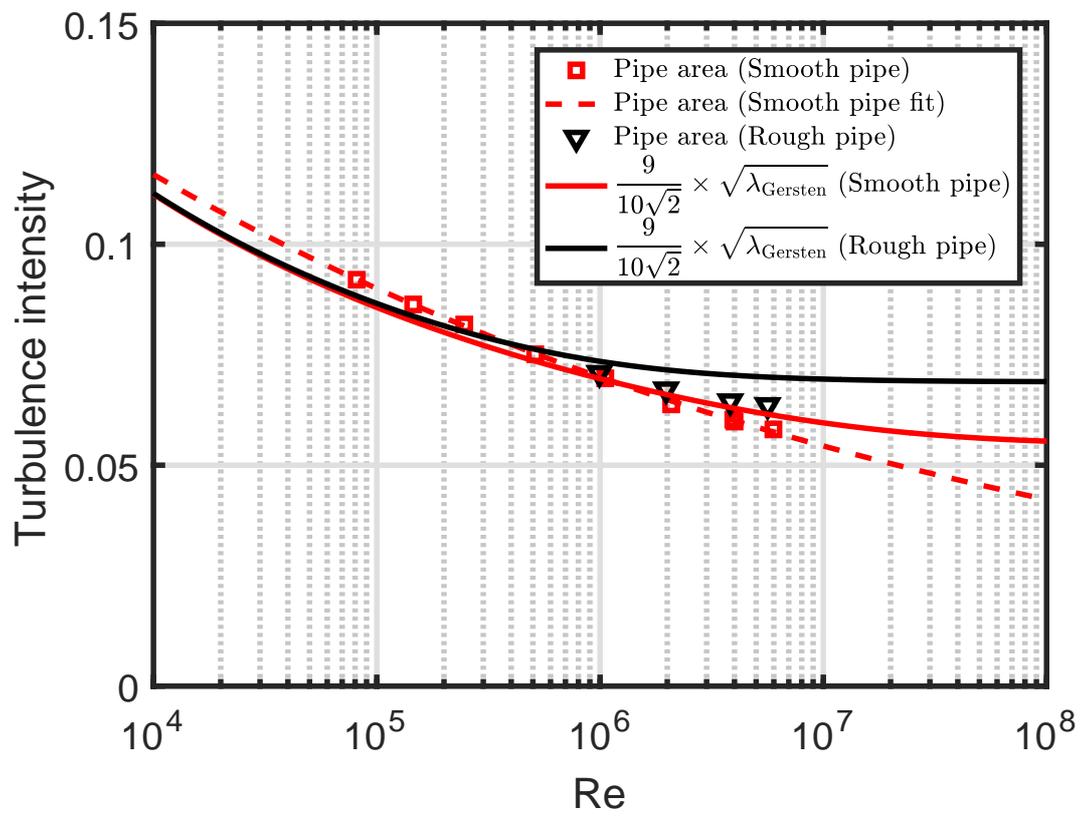}
\caption{Turbulence intensity for smooth and rough pipe flow. The approximation in Eq. (\ref{eq:ti_ff}) is included for comparison.}
\label{fig:NEW_comp_fric_ti}
\end{figure}

We show how well this approximation works in Fig. \ref{fig:NEW_comp_fric_ti}. Overall, the agreement is within 15\%.

\vspace{0.5cm}

We proceed to define the average kinetic energy of the turbulent velocity fluctuations $\langle E_{\rm kin,RMS} \rangle$ (per pipe volume $V$) as:

\begin{equation}
\begin{array}{ll}
\langle E_{\rm kin,RMS} \rangle/V & = \frac{1}{2} \rho \langle v_{\rm RMS} \rangle^2 \sim - \frac{81}{50} \times \left( \Delta P/L \right) D/4 \\ & = \frac{81}{50} \times \tau_w = \frac{81}{50} \times v_{\tau}^2 \rho,
\end{array}
\end{equation}

\noindent with $V=L \pi R^2$ so we have:

\begin{equation}
\begin{array}{ll}
\langle E_{\rm kin,RMS} \rangle & = \frac{1}{2} m \langle v_{\rm RMS} \rangle^2 \sim - \frac{81}{50} \times \left( \pi/2 \right) R^3 \Delta P \\ & = \frac{81}{50} \times \tau_w V = \frac{81}{50} \times v_{\tau}^2 m,
\label{eq:e_kin}
\end{array}
\end{equation}

\noindent where $m$ is the fluid mass. The pressure loss corresponds to an increase of the turbulent kinetic energy. The turbulent kinetic energy can also be expressed in terms of the mean flow velocity and the TI or the friction factor:

\begin{equation}
\langle v_{\rm RMS} \rangle^2 = v_m^2 I_{\rm Pipe~area}^2 \sim \frac{81}{200} \times v_m^2 \lambda
\end{equation}

\section{Discussion}
\label{sec:disc}

An overview of the general properties of turbulent velocity fluctuations can be found in \cite{marusic_a}.

\subsection{The attached eddy hypothesis}

Our quantification of the ratio $\langle v_{\rm RMS} \rangle/v_{\tau}$ as a constant can be placed in the context of the attached eddy hypothesis by Townsend \cite{townsend_a} \cite{marusic_b}. Our results are for quantities averaged over the pipe radius whereas the attached eddy hypothesis provides a local scaling with distance from the wall. By proposing an overlap region (see Fig. 1 in \cite{mckeon_b}) between the inner and outer scaling \cite{millikan_a}, it can be deduced that $\langle v_{\rm RMS} \rangle/v_{\tau}$ is a constant in this overlap region \cite{perry_a,perry_b}. Such an overlap region has been shown to exist in \cite{perry_a,hultmark_a}. The attached eddy hypothesis has provided the basis for theoretical work on e.g. the streamwise turbulent velocity fluctuations in flat-plate \cite{marusic_c} and pipe flow \cite{hultmark_c} boundary layers. Work on the law of the wake in wall turbulence also makes use of the attached eddy hypothesis \cite{krug_a}.

As a consistency check for our results, we can compare the constant $9/5$ in Eq. (\ref{eq:rms_fric_approx}) to the prediction by Townsend:

\begin{equation}
\frac{v_{\rm RMS,Townsend}(r)^2}{v_\tau^2} = B_1 -A_1 \ln \left( \frac{R-r}{r} \right),
\label{eq:town_pred}
\end{equation}

\noindent where fits have provided the constants $B_1 = 1.5$ and $A_1 = 1.25$. Here, $A_1$ is a universal constant whereas $B_1$ is not expected to be a constant for different wall-bounded flows \cite{marusic_d}. The constants are averages of fits presented in \cite{hultmark_b} to the smooth- and rough-wall Princeton Superpipe measurements. The Townsend-Perry constant $A_1$ was found to be 1.26 in \cite{marusic_d}. Performing the area averaging yields:

\begin{equation}
\frac{\langle v_{\rm RMS,Townsend} \rangle^2}{v_\tau^2} = B_1 + \frac{3}{2} \times A_1 = 3.38
\label{eq:town_ln}
\end{equation}

Our finding is:

\begin{equation}
\frac{\langle v_{\rm RMS} \rangle^2}{v_\tau^2} \sim \left( \frac{9}{5} \right)^2 = 3.24,
\end{equation}

\noindent which is within 5\% of the result in Eq. (\ref{eq:town_ln}). The reason that our result is smaller is that Eq. (\ref{eq:town_pred}) is overpredicting the turbulence level close to the wall and close to the pipe axis. Eq. (\ref{eq:town_pred}) as an upper bound has also been discussed in \cite{pullin_a}.

\subsection{The friction factor and turbulent velocity fluctuations}

The proportionality between the average kinetic energy of the turbulent velocity fluctuations and the friction velocity squared has been identified in \cite{yakhot_a} for $Re > 10^5$. This corresponds to our Eq. (\ref{eq:e_kin}).

A correspondence between the wall-normal Reynolds stress and the friction factor has been shown in \cite{orlandi_a}. Those results were found using direct numerical simulations. The main difference between the cases is that we use the streamwise Reynolds stress. However, for an eddy rotating in the streamwise direction, both a wall-normal and a streamwise component should exist which connects the two observations.

\subsection{The turbulence intensity and the diagnostic plot}

Other related work can be found beginning with \cite{alfredsson_a} where the diagnostic plot was introduced. In following publications a version of the diagnostic plot was brought forward where the local TI is plotted as a function of the local streamwise velocity normalised by the free stream velocity \cite{alfredsson_b,alfredsson_c,castro_a}. Eq. (3) in \cite{alfredsson_c} corresponds to our $I_{\rm Core}$, see Eq. (\ref{eq:TI_prof_equation}) in \ref{app:ti_prof_fits}.

\section{Conclusions}
\label{sec:conc}

We have compared TI profiles for smooth- and rough-wall pipe flow measurements made in the Princeton Superpipe.

The  change of the TI profile with increasing $Re$ from hydraulically smooth to fully rough flow exhibits propagation from the pipe wall to the pipe axis. The TIR at $r_n=0.99$ scales linearly with $Re$.

The scaling of TI with $Re$ - on the pipe axis and averaged over the pipe area - shows that the smooth- and rough-wall level deviates with increasing Reynolds number.

We find that $I_{\rm Pipe~area} \sim \frac{9}{10 \sqrt{2}} \times \sqrt{\lambda}$. This relationship can be useful to calculate the TI given a known $\lambda$, both for smooth and rough pipes. It follows that given a pressure loss in a pipe, the turbulent kinetic energy increase can be estimated.

\section*{Acknowledgement}

We thank Professor A.J.Smits for making the Superpipe data publicly available \cite{smits_a}.

\newpage

\appendix

\section{Fits to the turbulence intensity profile}
\label{app:ti_prof_fits}

\begin{figure}[htbp]
\includegraphics[width=7cm]{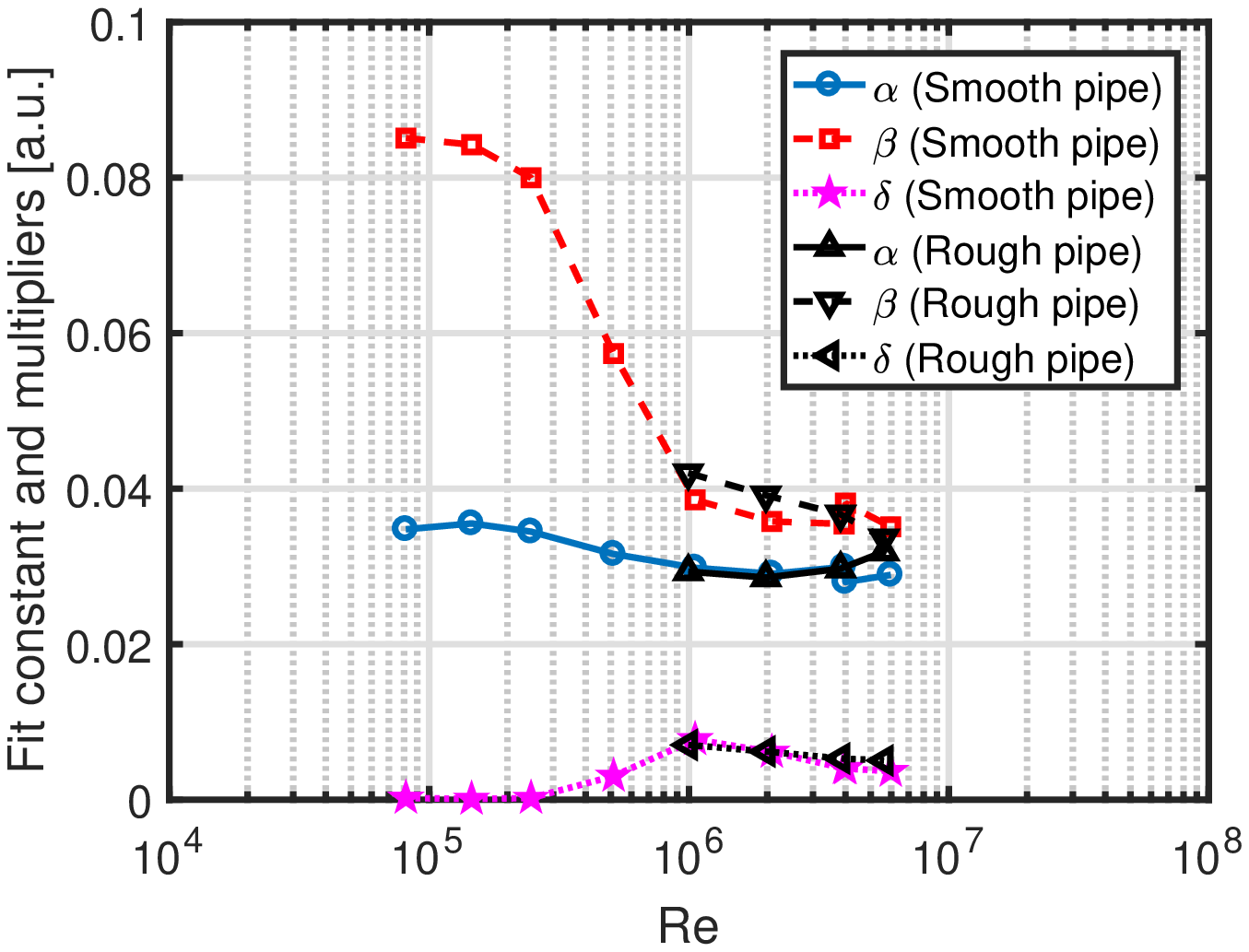}
\hspace{0.5cm}
\includegraphics[width=7cm]{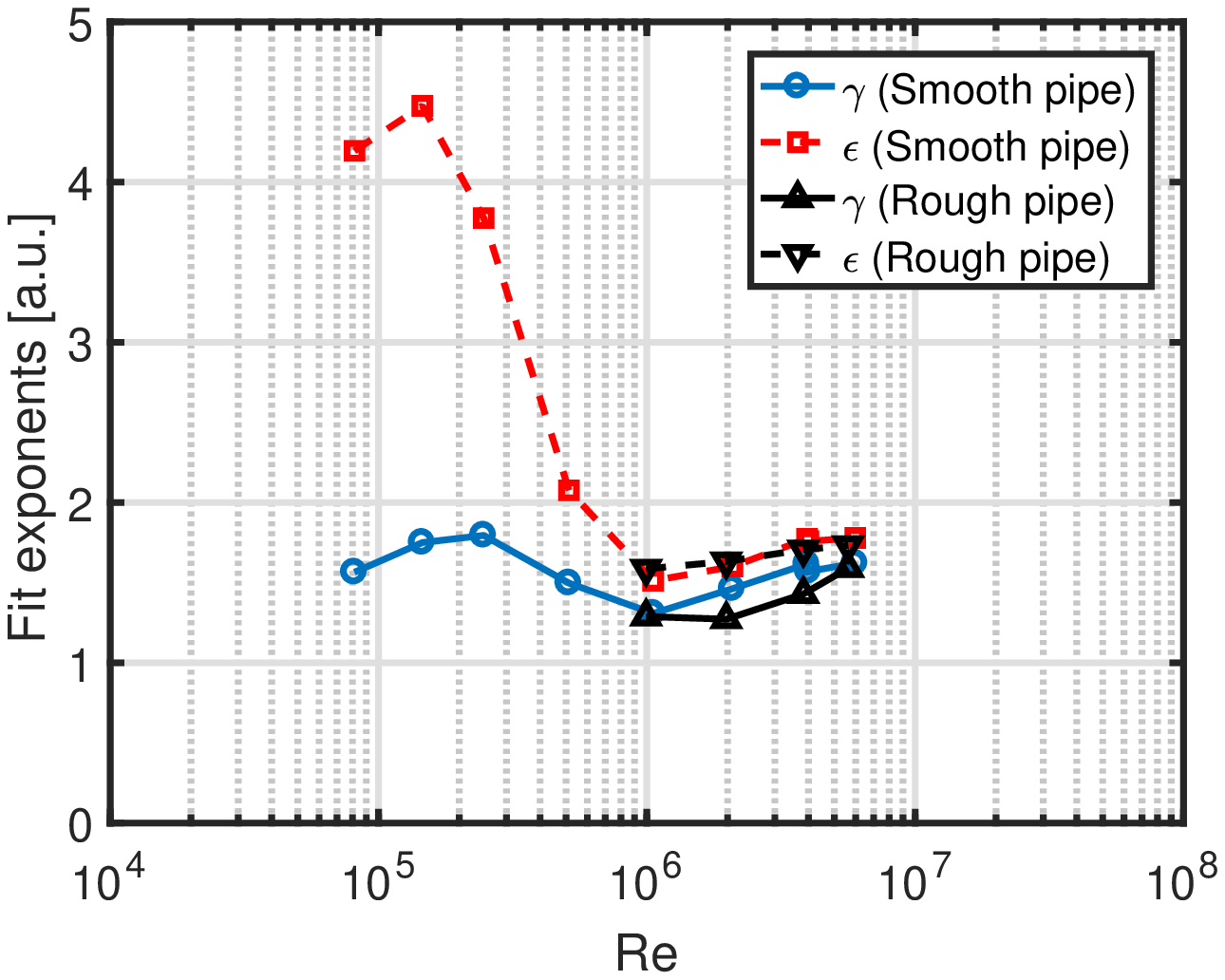}
\caption{Comparison of smooth- and rough-pipe fit parameters.}
\label{fig:smooth_rough_fits}
\end{figure}

As we have done for the smooth pipe measurements in \cite{russo_a}, we can also fit the rough pipe measurements to this function:

\begin{equation}
\begin{array}{ll}
I(r_n) & = I_{\rm Core}(r_n) + I_{\rm Wall}(r_n) \\ & = \left[ \alpha + \beta \times r_n^{\gamma} \right] + \left[ \delta \times |{\rm ln}(1-r_n)|^{\varepsilon} \right],
\end{array}
\label{eq:TI_prof_equation}
\end{equation}

\noindent where $\alpha$, $\beta$, $\gamma$, $\delta$ and $\varepsilon$ are fit parameters. A comparison of fit parameters found for the smooth- and rough-pipe measurements is shown in Fig. \ref{fig:smooth_rough_fits}. Overall, we can state that the fit parameters for the smooth and rough pipes are in a similar range for $10^6<Re<6 \times 10^6$.

\begin{figure}[htbp]
\includegraphics[width=7cm]{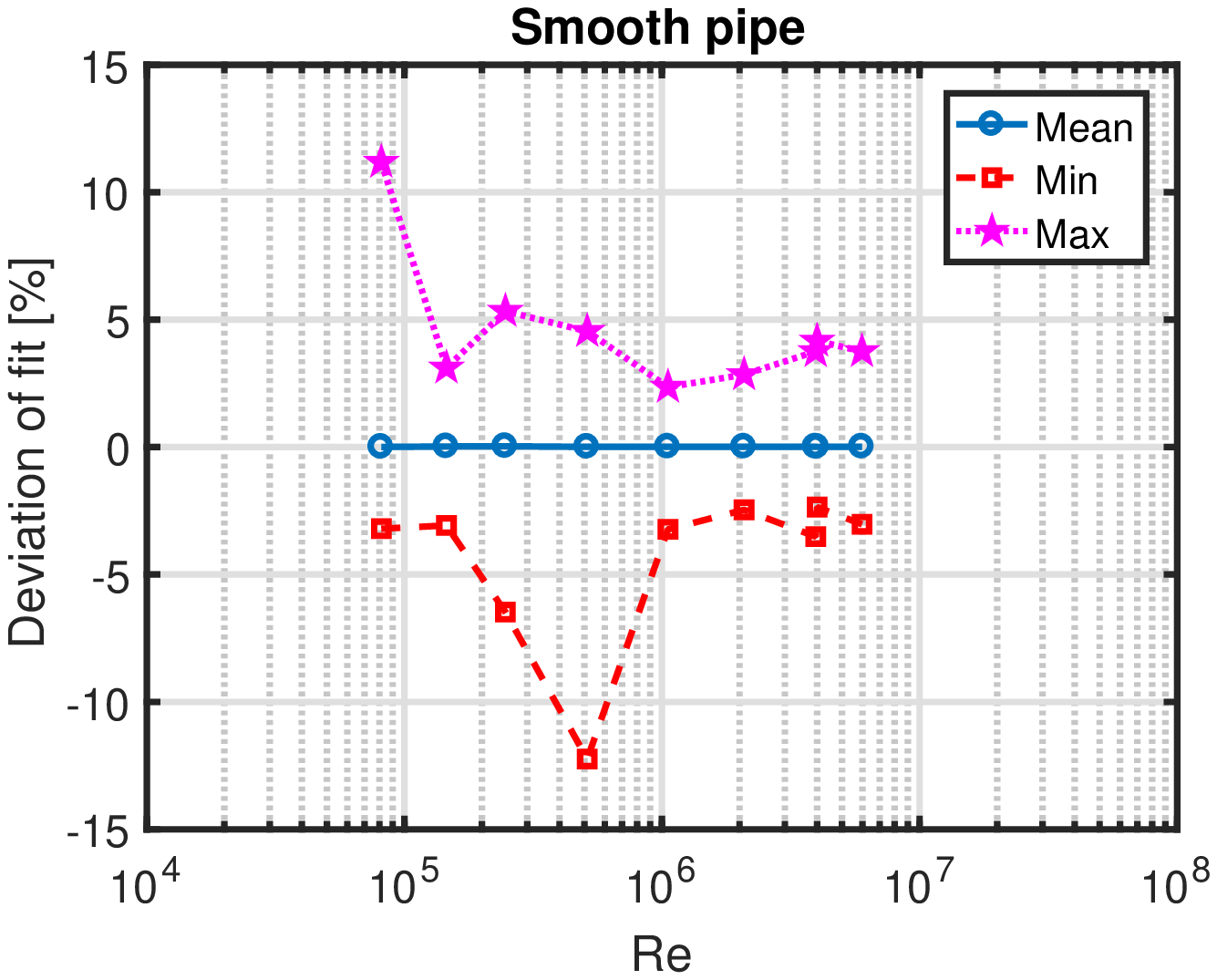}
\hspace{0.5cm}
\includegraphics[width=7cm]{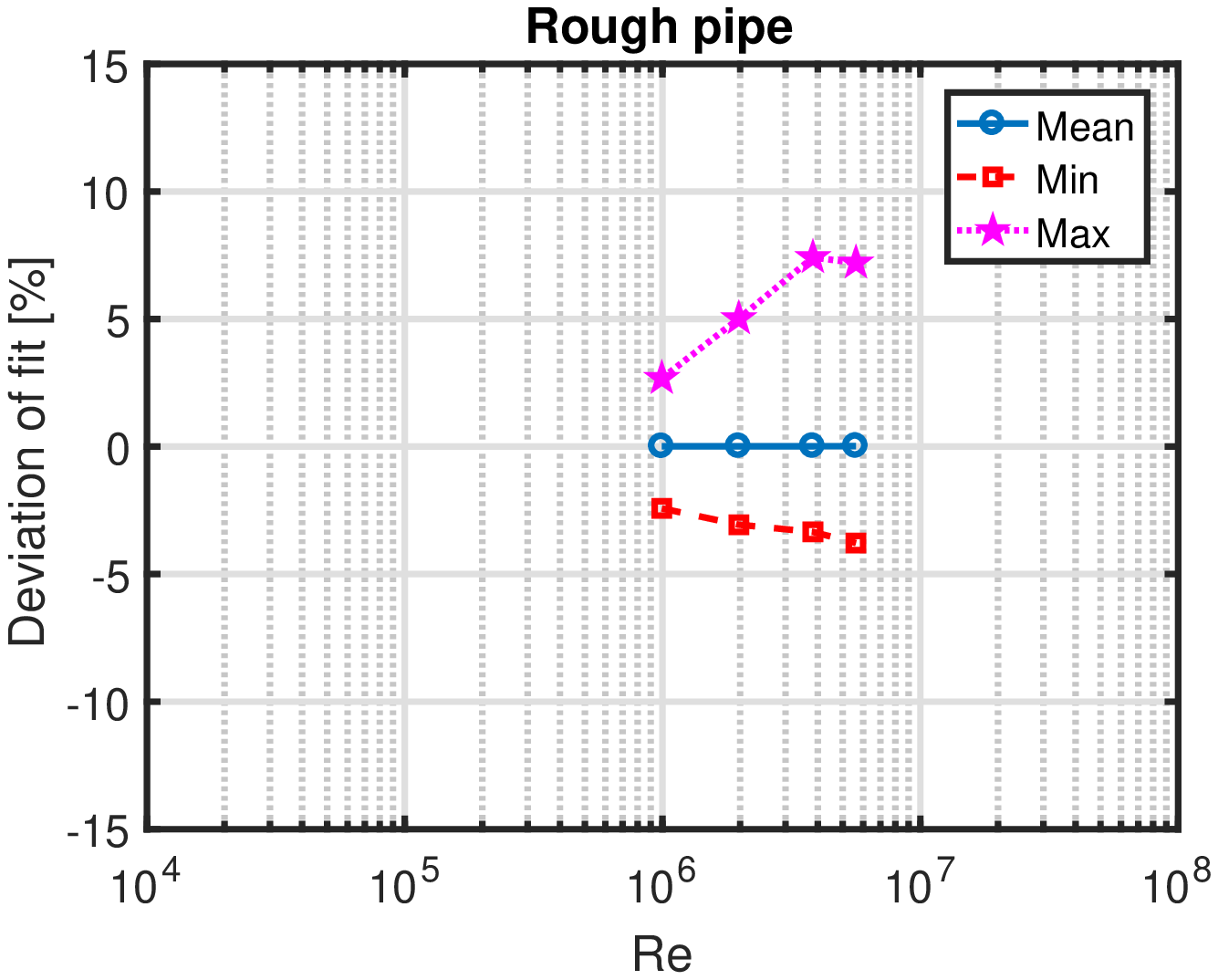}
\caption{Deviation of fits to measurements, left: Smooth pipe, right: Rough pipe.}
\label{fig:smooth_rough_deviations}
\end{figure}

The min/max deviation of the rough pipe fit from the measurements is below 10\%; see the comparison to the smooth wall fit min/max deviation in Fig. \ref{fig:smooth_rough_deviations}.

\begin{figure}[htbp]
\includegraphics[width=7cm]{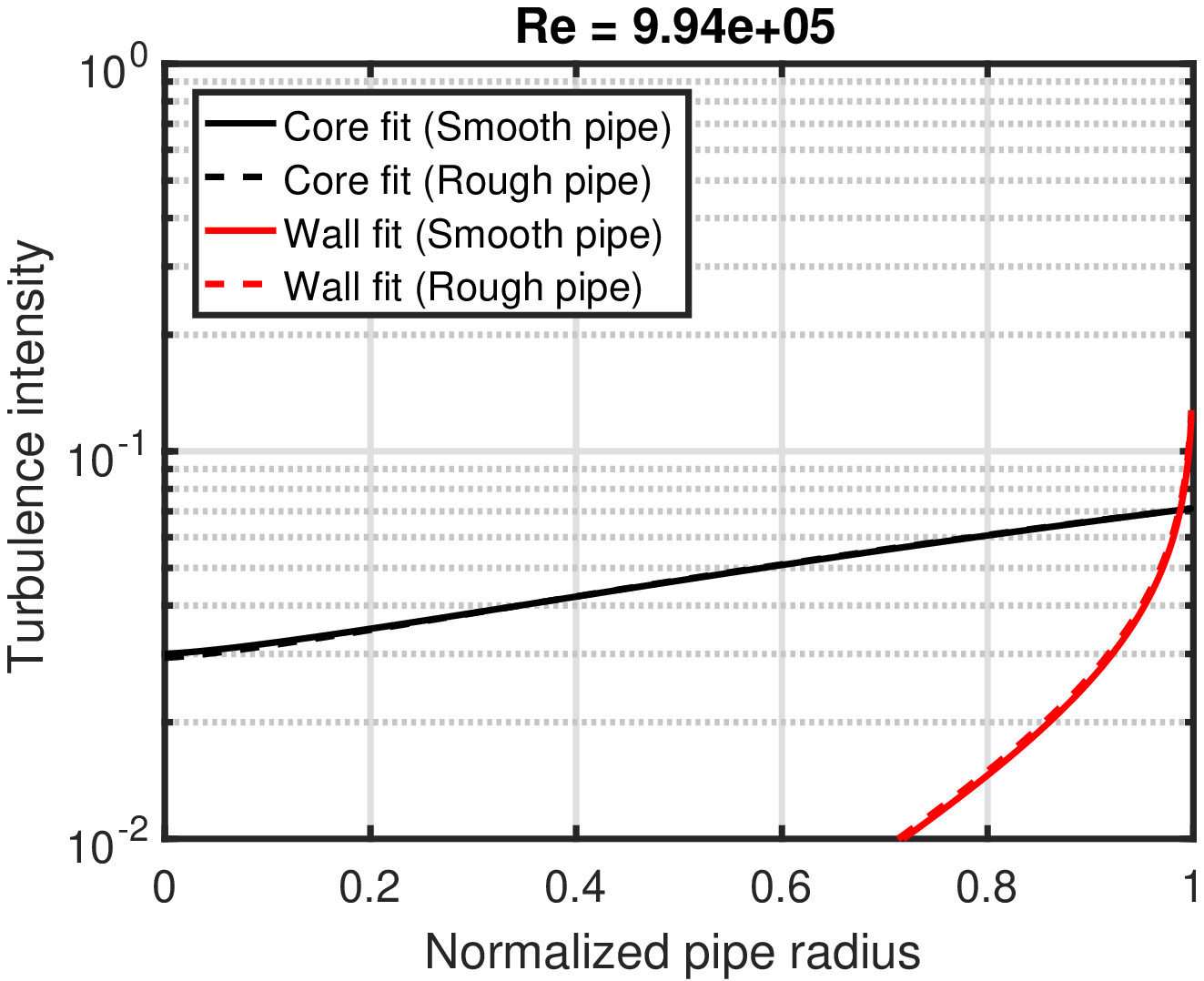}
\hspace{0.5cm}
\includegraphics[width=7cm]{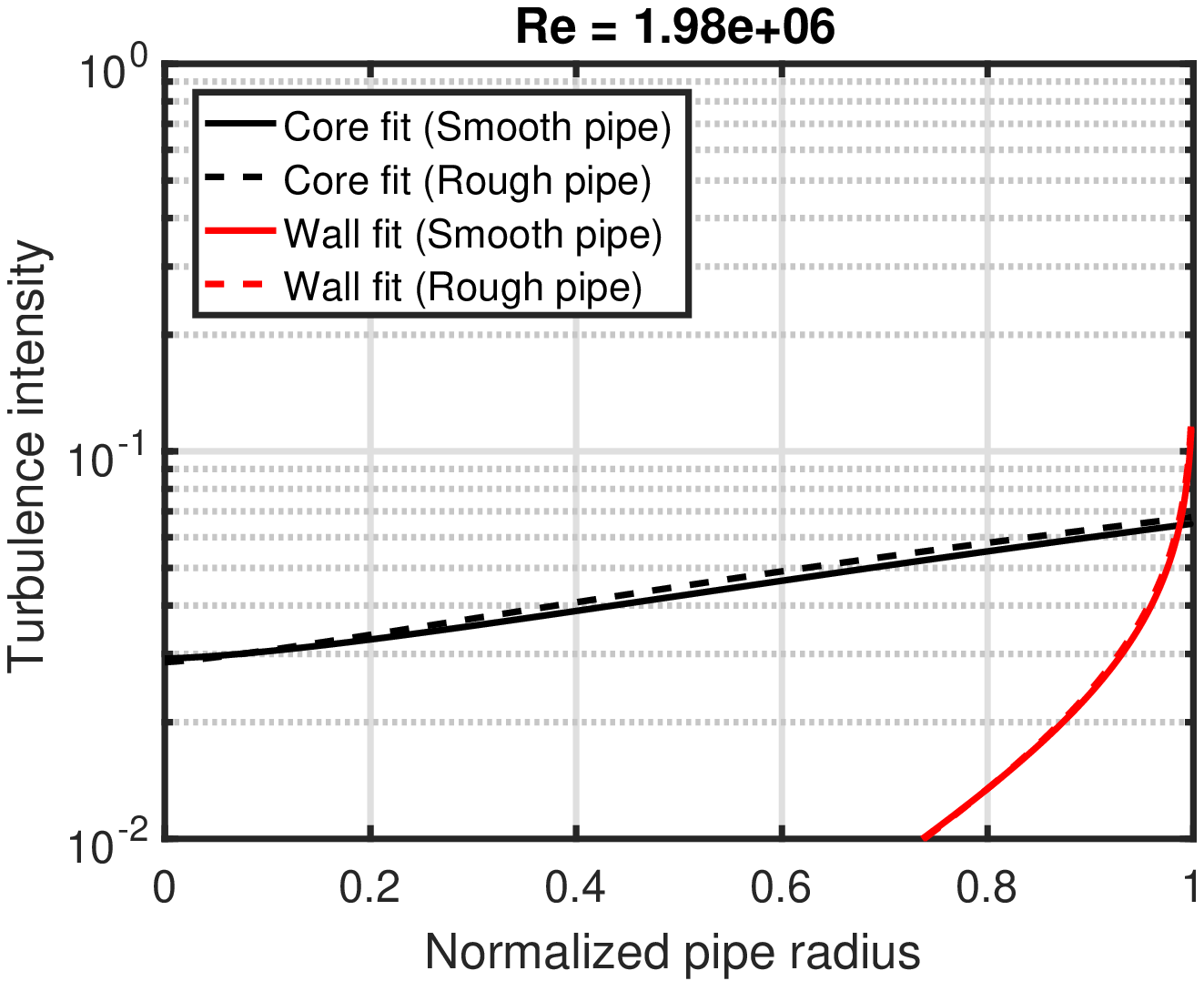}

\vspace{0.5cm}
\includegraphics[width=7cm]{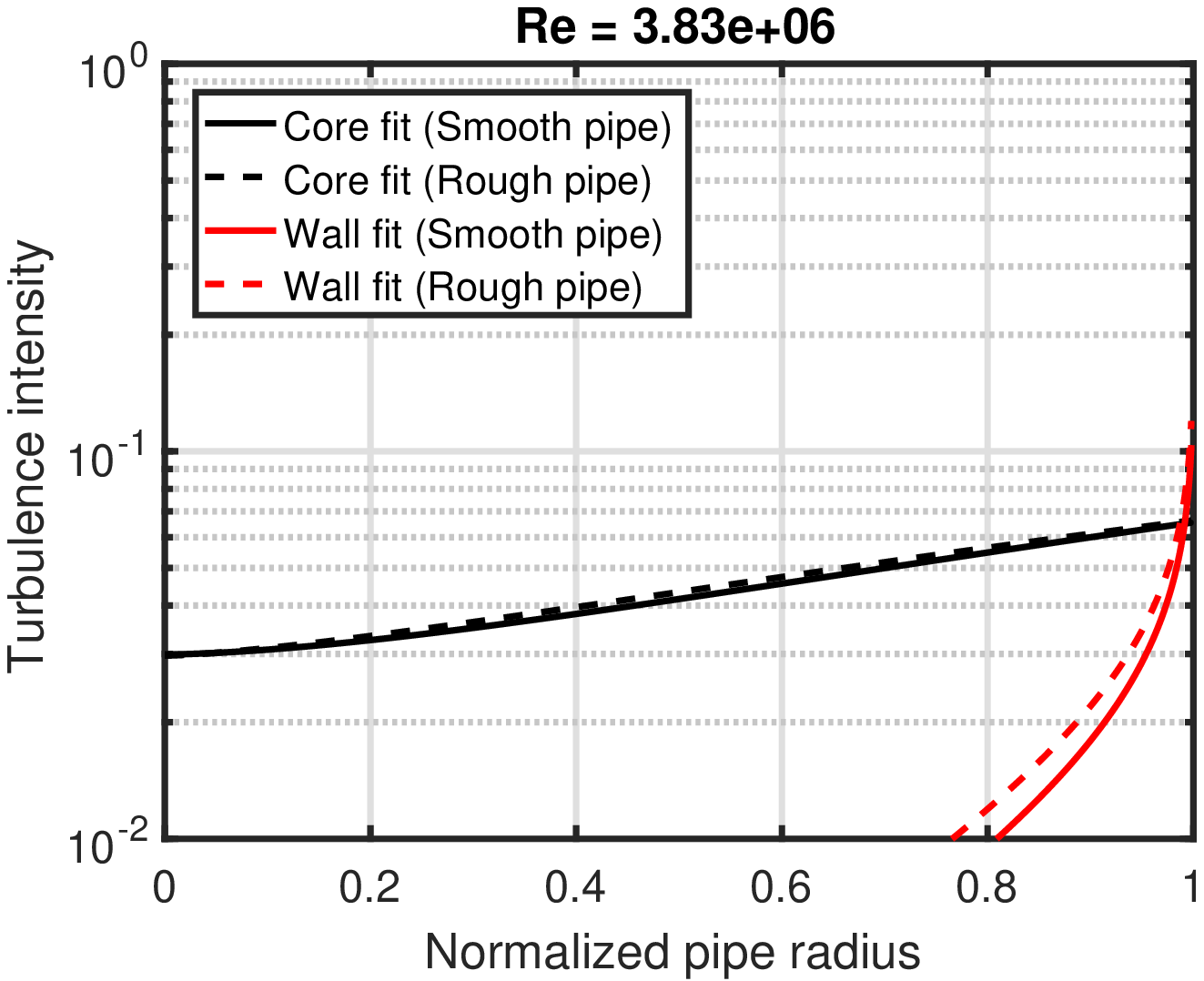}
\hspace{0.5cm}
\includegraphics[width=7cm]{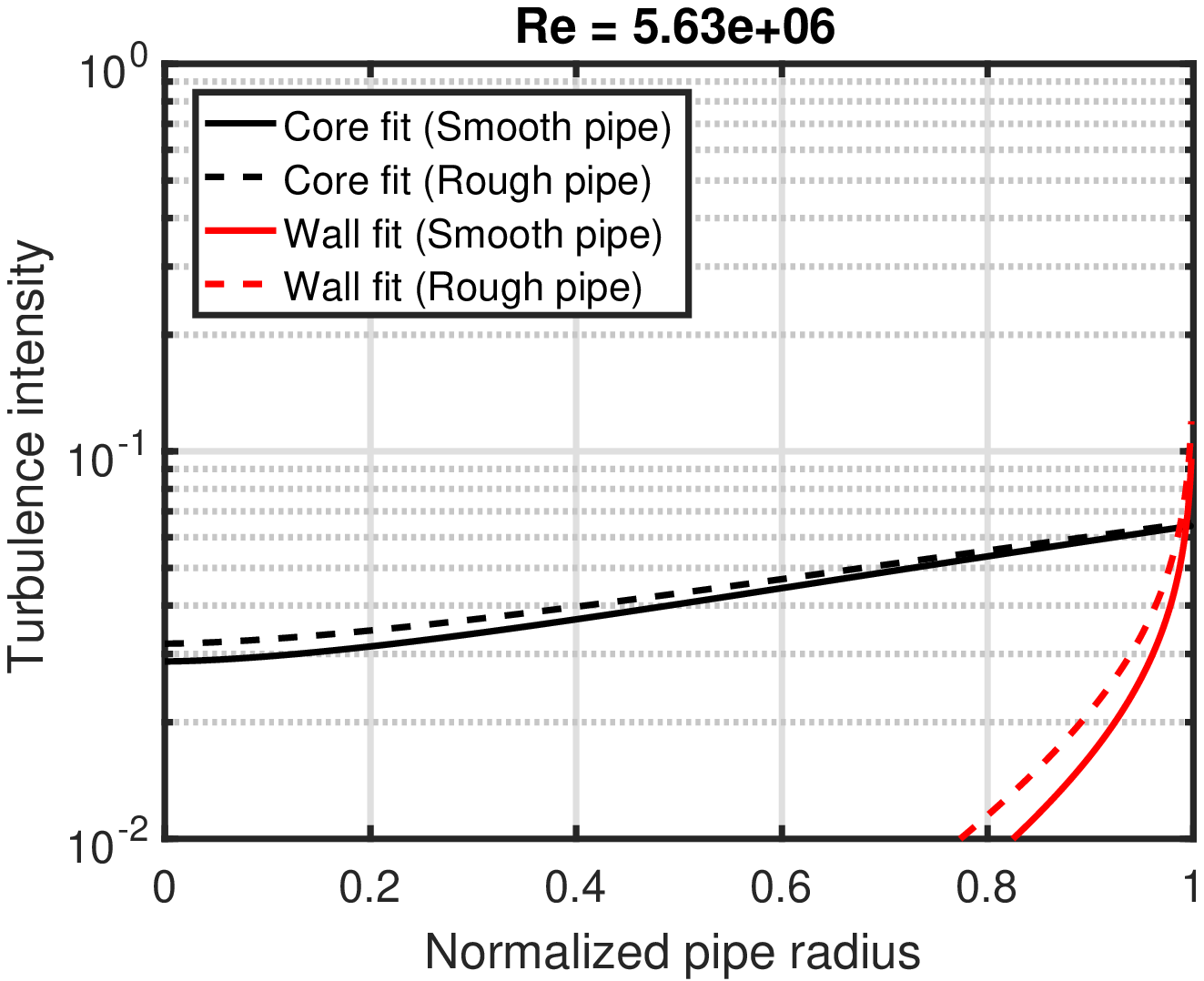}
\caption{Comparison of smooth and rough pipe core and wall fits.}
\label{fig:core_wall_fits}
\end{figure}

The core and wall fits for the smooth and rough pipe fits are compared in Fig. \ref{fig:core_wall_fits}. Both the core and wall TI increase for the largest $Re$.

\begin{figure}[htbp]
\includegraphics[width=14cm]{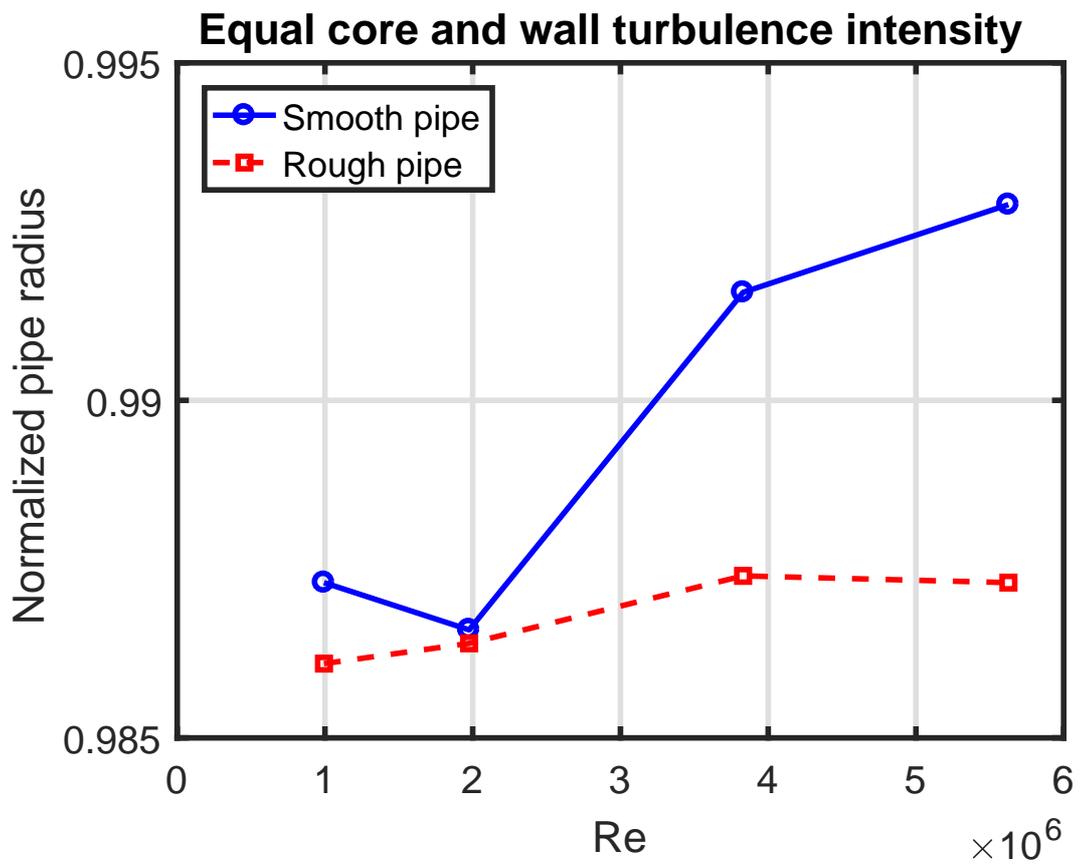}
\caption{Normalised pipe radius where the core and wall TI levels are equal.}
\label{fig:equal_ti_core_wall}
\end{figure}

The position where the core and wall TI levels are equal is shown in Fig. \ref{fig:equal_ti_core_wall}. This position does not change significantly for the rough pipe; however, the position does increase with $Re$ for the smooth pipe: This indicates that the wall term becomes less important relative to the core term.

\newpage

\section{Arithmetic mean definition of turbulence intensity averaged over the pipe area}
\label{app:alt_def_ti}

In the main paper, we have defined the TI over the pipe area in Eq. (\ref{eq:ti_area_AD}). In \cite{russo_a}, we used the arithmetic mean (AM) instead:

\begin{equation}
I_{\rm Pipe~area,~AM} = \frac{1}{R} \int_0^R \frac{v_{\rm RMS}(r)}{v(r)} {\rm d}r
\label{eq:ti_area_def}
\end{equation}

\begin{figure}[htbp]
\includegraphics[width=14cm]{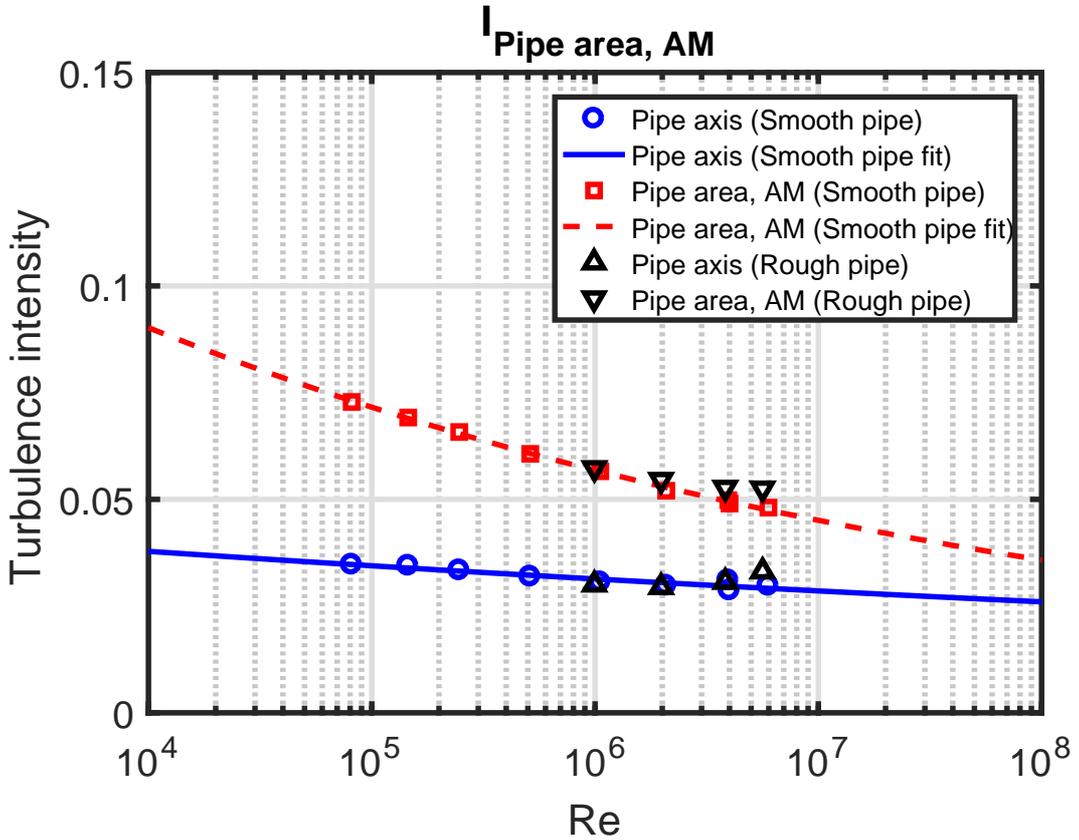}
\caption{Turbulence intensity for smooth and rough pipe flow. The AM is used for the pipe area TI.}
\label{fig:scal_smooth_rough_AD}
\end{figure}

The AM leads to a somewhat different pipe area scaling for the smooth pipe measurements which is illustrated in Fig. \ref{fig:scal_smooth_rough_AD}. Compare to Fig. \ref{fig:scal_smooth_rough}.

The scaling found in \cite{russo_a} using this definition is:

\begin{equation}
I_{\rm Smooth~pipe~area,~AM} = 0.227 \times Re^{-0.100}
\end{equation}

The AM scaling also has implications for the relationship with the Blasius friction factor scaling (Eq. (\ref{eq:blasius_ti_rel})):

\begin{equation}
\begin{array}{lcl}
I_{\rm Smooth~pipe~area,~AM} &=& 0.360 \times \lambda_{\rm Blasius}^{0.4} \\
\lambda_{\rm Blasius} &=&  12.89 \times I_{\rm Smooth~pipe~area,~AM}^{2.5}
\end{array}
\label{eq:blasius_ti_rel_AD}
\end{equation}

We can now define the AM version of the average velocity of the turbulent fluctuations:

\begin{equation}
\langle v_{\rm RMS} \rangle_{\rm AM} = v_m I_{\rm Pipe~area,~AM} = \frac{2}{R^3} \int_0^R v(r) r {\rm d}r \int_0^R \frac{v_{\rm RMS}(r)}{v(r)} {\rm d}r
\label{eq:avrg_v_rms_AD}
\end{equation}

The AM definition can be considered as a first order moment equation for $v_{\rm RMS}$, whereas the definition in Eq. (\ref{eq:avrg_v_rms}) is a second order moment equation.

\begin{figure}[htbp]
\includegraphics[width=14cm]{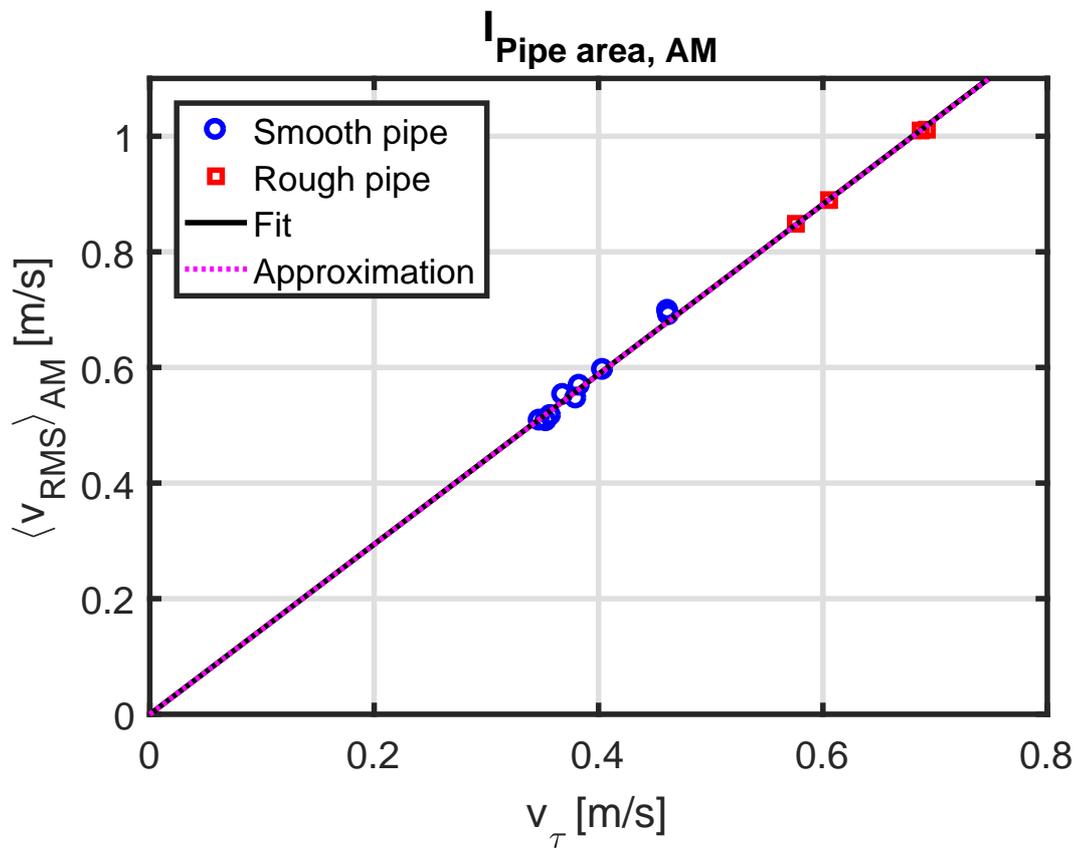}
\caption{Relationship between friction velocity and the AM average velocity of the turbulent fluctuations.}
\label{fig:fluc_fric_AD}
\end{figure}

Again, we find that the AM average turbulent velocity fluctuations are proportional to the friction velocity. However, the constant of proportionality is different than the one in Eq. (\ref{eq:rms_fric_fit}), see Fig. \ref{fig:fluc_fric_AD}. The AM case can be fitted as:

\begin{equation}
\langle v_{\rm RMS} \rangle_{\rm AM} = 1.4708 \times v_{\tau},
\label{eq:rms_fric_AD_rev}
\end{equation}

\noindent which we approximate as:

\begin{equation}
\langle v_{\rm RMS} \rangle_{\rm AM} \sim \sqrt{ \frac{2}{3} } \times \frac{9}{5} \times v_{\tau} \sim \sqrt{\frac{2}{3}} \times \langle v_{\rm RMS} \rangle
\label{eq:rms_fric_AD_approx}
\end{equation}

As we did in Section \ref{sec:disc}, we can perform the AM averaging of Eq. (\ref{eq:town_pred}) (also done in \cite{pullin_a}):

\begin{equation}
\frac{\langle v_{\rm RMS,Townsend} \rangle_{\rm AM}^2}{v_\tau^2} = B_1 + A_1 = 2.75,
\label{eq:town_am}
\end{equation}

\noindent where we find:

\begin{equation}
\frac{\langle v_{\rm RMS} \rangle_{\rm AM}^2}{v_\tau^2} \sim \frac{2}{3} \times \left( \frac{9}{5} \right)^2 = 2.16
\end{equation}

\newpage

\section*{References}
\label{sec:refs}

\end{document}